\documentclass[12pt,a4paper]{JHEP3} 

\JHEPspecialurl{http://jhep.sissa.it/JOURNAL/JHEP3.tar.gz}

\usepackage{amsmath}
\usepackage{graphicx}
\usepackage{latexsym}
\usepackage{amsfonts}
\usepackage{amssymb}
\usepackage{epsfig}

\newcommand\fverb{\setbox\fverbbox=\hbox\bgroup\verb}
\newcommand\fverbdo{\egroup\medskip\noindent%
            \fbox{\unhbox\fverbbox}\ }
\newcommand\fverbit{\egroup\item[\fbox{\unhbox\fverbbox}]}
\newbox\fverbbox

\newcommand{\drawsquare}[2]{\hbox{%
\rule{#2pt}{#1pt}\hskip-#2pt
\rule{#1pt}{#2pt}\hskip-#1pt
\rule[#1pt]{#1pt}{#2pt}}\rule[#1pt]{#2pt}{#2pt}\hskip-#2pt
\rule{#2pt}{#1pt}}

\newcommand{\fund}{~\raisebox{-.5pt}{\drawsquare{6.5}{0.4}}~}
\newcommand{\antifund}{~\overline{\raisebox{-.5pt}{\drawsquare{6.5}{0.4}}}~}

\newcommand{\antisym}{~\raisebox{-3.5pt}{\drawsquare{6.5}{0.4}}\hskip-6.9pt%
        \raisebox{3pt}{\drawsquare{6.5}{0.4}}~}

\newcommand{\f}{\frac}
\newcommand{\Tr}{{\rm Tr}}

\newcommand{\be}{\begin{equation}}
\newcommand{\ee}{\end{equation}}
\newcommand{\bea}{\begin{eqnarray}}
\newcommand{\eea}{\end{eqnarray}}

\newcommand{\OO}{\mathcal{O}}

\renewcommand{\t}{\tilde}
\newcommand{\mc}{\mathcal}
\newcommand{\tr}{\rm{tr}}

\title{New dynamics and dualities in supersymmetric chiral gauge theories}

\author{Nathaniel Craig$^{1,\,2}$, Rouven Essig$^3$ ,  Anson Hook$^3$, Gonzalo Torroba$^3$
\\
\vspace{0.2cm}
~\\
$^1$ School of Natural Sciences, \\
Institute for Advanced Study, Princeton, NJ 08540, USA \\
\vspace{0.2cm}

$^2$ Department of Physics and Astronomy \\
Rutgers University, Piscataway, NJ 08854, USA \\
\vspace{0.2cm}

$^3$ Physics Department and SLAC National Accelerator Laboratory \\
Stanford University, CA 94309, USA \\
\vspace{0.2cm}

}

\abstract{We analyze the phase structure of supersymmetric chiral gauge theories with gauge group $SU(N)$, an antisymmetric, and $F \le N+3$ flavors, in the presence of a cubic superpotential. When $F=N+3$ the theory flows to a superconformal fixed point in the infrared, and new dual descriptions of this theory are uncovered. The theory with odd $N$ admits a self-dual magnetic description. For general $N$, we find an infinite family of magnetic dual descriptions, characterized by arbitrarily large gauge groups and additional classical global symmetries that are truncated by nonperturbative effects. The infrared dynamics of these theories are analyzed using $a$-maximization, which supports the claim that all these theories flow to the same superconformal fixed point. A very rich phase structure is found when
 the number of flavors is reduced below $N+3$, including a new self-dual point, transitions from conformal to confining, and a nonperturbative instability for $F\le N$. We also give examples of chiral theories with antisymmetrics that have nonchiral duals.
}

\preprint{SLAC-PUB-14480 \\RUNHETC-2011-14 }

\begin{document}


\section{Introduction}\label{sec:intro}

A central problem in quantum field theory is to understand the phases of interacting gauge theories. While results on nonsupersymmetric theories are mostly numerical, theories with $\mc N=1$ supersymmetry (four supercharges) have holomorphic quantities which, in some cases, can be used to determine the vacuum structure. During the last decades, enormous progress on $\mc N=1$ theories has been made following the work of Seiberg~\cite{Seiberg:1994bz,Intriligator:1995au}. Seiberg duality has now become a key tool for analyzing strongly coupled effects.

Chiral gauge theories exhibit many fascinating phenomena, and have important theoretical and phenomenological applications. Starting from the pioneering works of~\cite{Affleck:1984uz,Affleck:1984xz}, dynamical supersymmetry breaking was found in chiral theories, providing one of the main motivations for their subsequent study. For reviews and references see~\cite{Giudice:1998bp,Poppitz:1998vd,Shadmi:1999jy}. Furthermore, intriguing dualities and nonperturbative effects have been found in these theories, like chiral-nonchiral dualities~\cite{Pouliot:1995zc,Pouliot:1996zh}, mixed phases~\cite{Terning:1997jj}, and new phase transitions between conformal and confining theories (as we shall also find in this work). Various other examples have been studied for instance in~\cite{Intriligator:1995ax,Brodie:1996xm}.

However, a general understanding of the infrared (IR) dynamics of chiral theories is still lacking, as is a systematic procedure to obtain dual theories.  Progress in obtaining dual theories was made by Berkooz~\cite{Berkooz:1995km}, who proposed to `deconfine' fields transforming in 2-index representations.  The deconfinement method will also play an important part in this paper.

In this work, we present new dualities and dynamical effects in chiral gauge theories with an $SU(N)$ gauge group and matter in the antisymmetric and (anti) fundamental representations. Related work on this class of models appears in~\cite{Terning:1997jj,Berkooz:1995km,Pouliot:1995me,Csaki:2004uj}. Cancellation of gauge anomalies restricts the matter content to 
\begin{center} 
\be\label{table:chiral-general}
\begin{tabular}{c|c|cc}
&$SU(N)$&$SU(F)$&$SU(N+F-4)$  \\
\hline
&&&\\[-12pt]
$Q$&$\fund$&$\antifund$&$1$  \\
$\t Q$&$\antifund$&$1$&$\fund$\\
$A$& $\antisym$ &$1$&$1$
\end{tabular}
\ee
\end{center}
We will focus on the case $F\le N+3$ and add a cubic superpotential $W \supset \t Q A \t Q$ for some of the quarks, described in \S \ref{sec:self-dual}. The first part of the work (\S \S \ref{sec:self-dual}--\ref{sec:amax}) analyzes the dynamics for $F=N+3$, while the phase structure when $F<N+3$ is studied in \S \ref{sec:mass}.
The case $F>N+3$ will be studied in~\cite{appear}.

First, in \S \ref{sec:self-dual} we argue that for $N$ odd and $F=N+3$ this theory has a dual magnetic description in terms of an $SU(N)$ gauge theory that includes additional mesons, baryons and cubic interactions.  This reveals that this theory, which features a nonzero superpotential, is self-dual (i.e.~the electric and magnetic descriptions have the same gauge group). This extends the results of~\cite{Csaki:1997cu,Karch:1997jp} on self-dual theories.

In \S \ref{sec:K} we propose that the above electric theory with $F=N+3$ (now with $N$ arbitrary) admits an \textit{infinite family} of magnetic duals with gauge groups $SU(N+K-1)$ and with matter charged under an additional classical $SU(K)$ global symmetry, where $K$ runs over all the integers with the same parity as $N$. This duality is quite striking in two respects. First, for fixed electric gauge group (fixed $N$), the 
dual theories have an arbitrarily large gauge group but all flow to the same fixed point in the IR. This signals a dramatic reduction in the number of degrees of freedom due to renormalization group (RG) effects, which will be understood using $a$-maximization~\cite{Intriligator:2003jj}. Furthermore, the magnetic theories have additional global symmetries in the ultraviolet (UV). For the duality to hold, these symmetries have to be truncated quantum-mechanically. This is also a general puzzle found in works on deconfinement (first noticed in~\cite{Luty:1996cg}), that will be addressed here.

We will explain in \S \ref{subsec:nonpert} how global symmetries can be reduced by nonperturbative effects. This happens because all the fields that transform nontrivially under $SU(K)$ are truncated from the chiral ring at the quantum level. Nonzero vacuum expectation values (vev's) for these fields produce nonperturbative superpotentials that cause the theory to develop runaway instabilities that remove all supersymmetric vacua.  The nonperturbative truncation of the classical chiral ring is a familiar effect that occurs also in vector-like supersymmetric QCD (SQCD) (reviewed in \S \ref{subsec:consistency}).  However, we believe that the reduction of global symmetries in the IR is a novel phenomenon, inherently related to the chiral nature of these theories.\footnote{The opposite effect, where the global symmetry group is accidentally enhanced at the IR fixed point, has been observed before~\cite{Leigh:1996ds,Distler:1996ub}.}

The low energy dynamics of the electric and magnetic descriptions with $F=N+3$ are analyzed in \S \ref{sec:amax}. At the origin of moduli space these theories all flow to the same superconformal fixed point. The exact anomalous dimensions are calculated using $a$-maximization, and a precise agreement between the electric and magnetic results is obtained. Furthermore, we find that the coefficient $a_{mag}$ of the $a$-anomaly in the family of magnetic theories is independent of $K$, and agrees with $a_{el}$. Since the central charge $a$ can be viewed as a measure of the number of degrees of freedom at the fixed point, this strongly suggests that the fields associated to the gauge group $SU(N+K-1)$ in the high energy theory are reduced by RG effects.

The phase structure of the theory with $F \le N+2$ flavors turns out to be extremely rich, and this is the subject of \S \ref{sec:mass}. Depending on the mass deformation in the $F=N+3$ theory, the flow to $F=N+2$ results in either a self-dual conformal fixed point or an s-confining theory. Next, starting from either of these theories and adding one more mass term, we find that the theory with $F=N+1$ confines with chiral symmetry breaking. When $F\le N$ the theory develops a runaway instability caused by nonperturbative effects.  
A product gauge theory that interpolates between the electic and magnetic theories with $F=N+2$ is 
discussed in Appendix \S \ref{sec:prodflow}.  

Generalizations and applications of the previous results are contained in \S \ref{sec:Kpr}. We present an infinite family of electric theories with gauge group $SU(N+K-1)$ and a global symmetry group containing  $SU(K)$ that are dual to an infinite family of magnetic theories with gauge group $SU(N+K'-1)$ and a global symmetry factor $SU(K')$. A particular version of this gives a nonchiral dual to a chiral theory with antisymmetrics.

Finally, \S \ref{sec:concl} summarizes our results and suggests some future directions. 
\section{A self-dual chiral theory}\label{sec:self-dual}

In this section we describe the simplest duality in the chiral theory with $F=N+3$, exhibiting a self-dual magnetic description.

\subsection{The electric theory}\label{subsec:odd-electric}

Let us describe the electric theory in detail. The matter content is given by
\begin{center}
\be\label{table:chiral-self}
\begin{tabular}{c|c|cc}
&$SU(N)$&$Sp(2N-2)$&$SU(N+3)$  \\
\hline
&&&\\[-12pt]
$Q$&$\fund$&$1$&$\antifund$  \\
$\t Q$&$\antifund$&$ \fund$ &$1$ \\
$\t P$&$\antifund$&$1$&$1$  \\
$A$&$\antisym$&$1$&$1$  \\
\end{tabular}
\ee
\end{center}
and the superpotential is
\be\label{eq:Wchiral1}
W_\text{el}=  \t Q A \t Q\,.
\ee
The gauge symmetry is $SU(N)$ and the nonabelian global symmetries are $Sp(2N-2) \times SU(N+3)$.\footnote{Our convention is $Sp(2) \sim SU(2)$.} Equation (\ref{eq:Wchiral1}) is the most general renormalizable superpotential compatible with the symmetries. This superpotential plays a crucial role in the dynamics of the theory. This section deals with odd $N$, while the case of even $N$ will be discussed in \S \ref{sec:K}.

Comparing with (\ref{table:chiral-general}), we have set $F=N+3$, and $2N-2$ quarks $\t Q$ have been coupled to the antisymmetric through (\ref{eq:Wchiral1}). We have not found a simple magnetic dual if the number of quarks $\t Q$ interacting with $A$ is different than $2N-2$; in such cases the dual description has a product gauge group. These cases -- and the model with an arbitrary number of total flavors -- will be studied in~\cite{appear}.

The gauge invariant operators are the mesons
\be\label{eq:mesons}
Q \t Q\;,\;Q \t P\;,\;\t Q A \t Q\;,\;\t Q A \t P\,,
\ee
and baryons
\be\label{eq:baryons}
\t Q^{N}\,,\;\t Q^{N-1} \t P\,,\;Q^k A^{(N-k)/2}\,,\; k=1,\,3,\,\ldots\,,N\,.
\ee
The invariants $\t Q A \t Q$, $\t Q A \t P$, $\t Q^N$ and $\t Q^{N-1} \t P$ are lifted by the superpotential (\ref{eq:Wchiral1}).

\subsection{The magnetic dual theory}\label{subsec:odd-magnetic}

In order to understand the possible magnetic duals of this theory, we first discuss the case $N=3$, for which the antisymmetric becomes an antifundamental. The electric theory is $SU(3)$ with 6 flavors $Q$ and $\bar Q=(\t Q, \t P, A)$, deformed by a baryon operator $W_\text{el}= \bar B$, where $\bar B= \t Q A \t Q$. We have combined all the antifundamentals into a single vector of quarks $\bar Q$, and recall that the baryons are
\be
\bar B_{i_1i_2 i_3}= \epsilon^{\alpha_1 \alpha_2 \alpha_3} \bar Q_{\alpha_1 i_1}\bar Q_{\alpha_2 i_2}\bar Q_{\alpha_3 i_3}\,.
\ee

The Seiberg dual is an $SU(3)$ theory with magnetic quarks $q$ and $\bar q=(\t q, \t p, a)$ and extra mesons and interactions,
\be
W_\text{mag}= \bar b+ q (Q\t Q) \t q + q (Q \t P) \t p + q (Q A) a\,,
\ee
where $(\phi_i\phi_j)$ denotes the composite meson associated to the electric fields $\phi_i \phi_j$, and $\bar b$ is the magnetic baryon that maps to $\bar B$ according to~\cite{Intriligator:1995au}
\be
\bar b_{i_1 i_2 i_3}=\frac{1}{3!} \sqrt{\frac{\Lambda^3}{\mu^3}} \epsilon_{i_1 \ldots i_6} \epsilon^{\alpha_1 \alpha_2 \alpha_3} \bar q_{\alpha_1}^{i_4}\bar q_{\alpha_2}^{i_5}\bar q_{\alpha_3}^{i_5}\,,
\ee
where $\Lambda$ is the dynamical scale of the electic theory and $\mu$ is needed to match 
dimensionful quantities between the magnetic and electric theories.  
Further details of Seiberg duality in the presence of baryon deformations may be found in~\cite{Leigh:1996ds}.

Guided by this, we propose that the electric theory (\ref{table:chiral-self}) with general odd $N$ has a dual description given by
\begin{center}
\be\label{table:chiral-self3}
\begin{tabular}{c|c|cc}
&$SU(N)$&$Sp(2N-2)$&$SU(N+3)$  \\
\hline
&&&\\[-12pt]
$q$&$\fund$&$1$&$ \fund$  \\
$\t q$&$\antifund$&$\fund$ &$1$ \\
$\t p$&$\antifund$&$ 1$ &$1$ \\
$a$&$\antisym$&$ 1$ &$1$ \\
$M_1$&$1$&$\fund$&$ \antifund$ \\
$M_2$&$1$&$1$&$ \antifund$ \\
$s$&$1$&$1$&$ \antifund$
\end{tabular}
\ee
\end{center}
with superpotential interactions\footnote{In this work, we will set the superpotential coefficients to one by field redefinitions.}
\be\label{eq:oddWmag}
W_\text{mag}=q M_1 \t q+\t q a \t q +  q s \t p+ q a^{(N-1)/2}M_2\,.
\ee
The magnetic superpotential contains a term $q a^{(N-1)/2}M_2$ that is nonrenormalizable at high energies. This term is needed for the consistency of the duality. The analysis in \S \ref{sec:amax} of the IR fixed point will show that this operator can be irrelevant or dangerously irrelevant depending on the value of $N$.
 
We will provide various tests for this proposal shortly, and argue that both theories flow to the same superconformal fixed point. The appearance of the baryonic superpotential deformation and other aspects of this duality will be explained using the deconfinement method, which we will explain in \S \ref{subsec:productK}.

The magnetic theory has the same gauge group as the electric theory, thus providing a self-dual description. Notice that the contribution of matter to the beta function is also equivalent to the SQCD self-dual point $N_f=2N$, although in our case there are additional singlets and interactions. Self-dual chiral theories have been studied for instance in~\cite{Csaki:1997cu,Karch:1997jp}, in electric theories without a superpotential. Our model provides an example of a self-dual theory based on a marginal superpotential (\ref{eq:Wchiral1}). We will also present other self-dual theories in \S\S \ref{subsec:tQ} and \ref{sec:Kpr}.

The gauge invariants of the magnetic theory can be constructed as in (\ref{eq:mesons}) and (\ref{eq:baryons}). The F-term conditions lift many of these combinations, and we are left with the singlets $M_1$, $M_2$ and $s$, and the baryons
\be
q^{j+3} a^{(N-j-3)/2}\,,\; j\le N-3\;,\;j \in 2 \mathbb{Z}_{\ge 0}\,.
\ee
The superpotential (\ref{eq:oddWmag}) also introduces additional constraints along the different branches of moduli space.

\subsection{Consistency checks of the duality}\label{subsec:consistency}

We now present various consistency tests on the proposed dual pairs (\ref{table:chiral-self}) and (\ref{table:chiral-self3}).

Let us start by matching the global symmetries and chiral rings of both theories. The anomaly free abelian symmetries of the electric theory with superpotential (\ref{eq:Wchiral1}) can be parametrized 
by\footnote{\label{foot:conditions}
The vanishing of the $U(1) SU(N)^2$ and $U(1)_R SU(N)^2$ anomalies requires
$T(G) + \sum_i T(r_i) (R_i - 1) = 0$ and $\sum_i T(r_i) \mc{Q}_i = 0$,
where $T(r)$ is defined via $\tr (t_r^a t_r^b) = T(r) \delta^{ab}$.  
Here $T(G_{SU(N)}) = N$, $T(\fund) = T(\antifund) = \f{1}{2}$ and
$T(\antisym) = \frac{N-2}{2}$. 
We also require that the superpotential terms  have $R$-charge 2 and vanishing charge $\mc Q$.
}
\begin{center}
\be\label{table:U1-electric}
\begin{tabular}{c|cc}
&$U(1)$&$U(1)_R$  \\
\hline
&&\\[-12pt]
$Q$&$\mc Q_Q$&$R_Q$  \\
$\t Q$&$\mc Q_{\t Q}$&$R_{\t Q}$ \\
$\t P$&$-(N+3) \mc Q_Q-2 \mc Q_{\t Q}$&$4-(N+3) R_Q-2 R_{\t Q}$  \\
$A$&$-2\mc Q_{\t Q}$&$2-2R_{\t Q}$ \\
\end{tabular}
\ee
\end{center}
where $\mc Q_Q$, $\mc Q_{\t Q}$, $R_Q$ and $R_{\t Q}$ represent arbitrary charge assignments. 
Note that the $R$-symmetry is not unique. In the magnetic theory of \S \ref{subsec:odd-magnetic} the corresponding charges read
\begin{center}
\be\label{table:U1-magnetic}
\begin{tabular}{c|cc}
&$U(1)$&$U(1)_R$  \\
\hline
&&\\[-12pt]
$q$&$\frac{1}{N}\left(3 \mc Q_Q-(N-3) Q_{\t Q} \right)$&$\frac{1}{N}\left(3R_Q+(N-3)(1-R_{\t Q}) \right)$\\
$\t q$&$-\frac{1}{N}\left((N+3) \mc Q_Q+3 \mc Q_{\t Q} \right)$&$\frac{1}{N}\left((N+3)(1- R_Q)-3 R_{\t Q} \right)$ \\
&&\\[-14pt]
$\t p$&$\frac{1}{N}\left(- (N+3) \mc Q_Q+(N^2-3)\mc Q_{\t Q}\right)$&$\frac{-(N+1)(N-3)-(N+3) R_Q+(N^2-3) R_{\t Q}}{N}$ \\
$a$&$\frac{2}{N}\left((N+3) \mc Q_Q+3 \mc Q_{\t Q} \right)$&$\frac{2}{N}\left((N+3) R_Q+3 R_{\t Q}-3\right)$ \\
$M_1$&$\mc Q_Q + \mc Q_{\t Q}$&$R_Q+R_{\t Q}$  \\
$M_2$&$-(N+2) \mc Q_Q-2 \mc Q_{\t Q}$&$4-(N+2) R_Q-2 R_{\t Q}$  \\
$s$&$\mc Q_Q-(N-1)\mc Q_{\t Q}$&$R_Q+(N-1)(1- R_{\t Q})$ \\
\end{tabular}
\ee
\end{center}
The 't Hooft anomaly matching conditions are satisfied for these symmetries.

Taking into account the anomaly-free global symmetries, the mapping of the chiral rings of the electric and magnetic theories is
\bea\label{eq:match-self}
Q \t Q &\leftrightarrow& M_1\nonumber\\
Q \t P &\leftrightarrow& M_2\nonumber\\
Q^{N-j} A^{j/2} &\leftrightarrow& q^{j+3} a^{(N-j-3)/2} \;,\; j \le N-3 \;,\; j \in 2 \mathbb{Z}_{\ge0} \nonumber \\
Q A^{(N-1)/2} &\leftrightarrow& s\,.
\eea
Notice that the elementary singlet $s$ in the magnetic description maps to a baryon of the electric theory.

In this correspondence, the agreement of ranks in the electric and magnetic descriptions needs to be explained. For instance, while $Q \t Q$ has rank $N$, the elementary meson $M_1$ has classical rank $min(2N-2,N+3)$. It is useful to recall that in $SU(N_c)$ SQCD with $N_f$ quarks $(Q, \t Q)$, a similar situation arises in the mapping $Q \t Q \leftrightarrow M$, where $M$ consists of the $N_f^2$ singlets in the magnetic theory and has classical rank $\le N_f$ instead of $\le N_c$, which is the rank of $Q\t Q$.  
The resolution in that case involves nonperturbative effects (e.g.~\cite{Intriligator:1995au}), 
which enforce the constraint rank$(M)\le N_c$.  Due to the superpotential $W_{mag}= q M \t q$ (where $q$ and $\t q$ are the magnetic quarks), a nonzero vev for $M$ of rank$(M)$ gives mass to rank$(M)$ flavors, but leaves $N_f-{\rm rank}(M)$ massless flavors.  Since the dual gauge group is SU($N_f-N_c$), the theory develops an Affleck-Dine-Seiberg (ADS) runaway for $N_f-{\rm rank}(M) < N_f-N_c$, 
i.e.~${\rm rank}(M) > N_c$, which destroys the vacuum.  Therefore, these extra components in $M$ are dynamically truncated from the chiral ring.

In our case, turning on $M_1$ with rank larger than $N$ also leads to an ADS-like superpotential with no supersymmetric vacua.  For instance, if $M_1$ has rank $N+1$ and assuming $N\ge 5$, the low energy theory has $N_f=2$ vector-like massless flavors (two $q$'s and two $\t q$'s) with a dynamical superpotential~\cite{Poppitz:1995fh}
\bea
\label{eq:NP1}
W_\text{dyn} = \frac{\Lambda_L^{2N+1} }{(q \t q) (q a^{\frac{N-1}{2}}) (\t q a \t q)^{\frac{N-3}{2}}}\,.
\eea
Here the low energy scale $\Lambda_L$ is related to the dynamical scale of the magnetic dual by $\Lambda_L^{2N+1}\sim \langle M_1^{N+1} \rangle\Lambda_\text{mag}^N $. There are no supersymmetric vacua for (\ref{eq:oddWmag}) plus (\ref{eq:NP1}), so a meson $M_1$ of rank larger than $N$ is blocked from the chiral ring.  (For the case $N=3$ discussed in \S \ref{subsec:odd-magnetic}, the resolution is the same as in the previous paragraph.)

Below we will present two additional tests for the duality. In \S \ref{subsec:productK} (setting $K=1$ there) we exhibit a product gauge group theory $SU(N) \times Sp(N-3)$ that flows to the electric description if $\Lambda_{Sp(N-3)} \gg \Lambda_{SU(N)}$, while for $\Lambda_{SU(N)} \gg \Lambda_{Sp(N-3)}$ the IR fixed point corresponds to the magnetic theory. We can then interpolate between the electric and magnetic descriptions by varying the holomorphic ratio $\Lambda_{SU(N)}/\Lambda_{Sp(N-3)}$, which should not 
lead to any phase transitions. This provides further strong evidence that both theories have the same phase structure and dynamics in the IR.

Furthermore, in \S \ref{sec:amax} we analyze the IR fixed point using $a$-maximization and find a precise agreement between the electric and magnetic predictions. This includes the exact anomalous dimensions and values for the superconformal $a$-function.

\section{An infinite family of dual theories}\label{sec:K}

In the previous section we studied the chiral gauge theory (\ref{table:chiral-self}) for $N$ odd and argued that this theory is self-dual, with a magnetic description given by (\ref{table:chiral-self3}). Now we consider the same theory for arbitrary $N$ and propose a new set of dualities. We will establish that there exists an infinite family of chiral theories with different gauge group rank and perturbative global symmetries, all of which flow to the same fixed point.

We will argue that the electric theory (\ref{table:chiral-self}),
\begin{center}
\be\label{table:repeat1}
\begin{tabular}{c|c|cc}
&$SU(N)$&$Sp(2N-2)$&$SU(N+3)$  \\
\hline
&&&\\[-12pt]
$Q$&$\fund$&$1$&$\antifund$  \\
$\t Q$&$\antifund$&$ \fund$ &$1$ \\
$\t P$&$\antifund$&$1$&$1$  \\
$A$&$\antisym$&$1$&$1$  \\
\end{tabular}
\ee
\end{center}
with superpotential
\be
W_\text{el}= \t Q A \t Q\,,
\ee
has an \textit{infinite family} of dual descriptions with matter content
\begin{center}
\be\label{table:dualK}
\begin{tabular}{c|c|ccc}
&$SU(N+K-1)$&$SU(K)$&$Sp(2N-2)$&$SU(N+3)$  \\
\hline
&&&\\[-12pt]
$q$&$\fund$&$1$&$1$&$ \fund$  \\
$\t q$&$\antifund$&$1$&$\fund$ &$1$ \\
$\t p$&$\antifund$&$\fund$&$ 1$ &$1$ \\
$a$&$\antisym$&$ 1$&$ 1$ &$1$ \\
$M_1$&$1$&$ 1$&$\fund$&$ \antifund$  \\
$M_2$&$1$&$ 1$&$1$&$ \antifund$  \\
$s_1$&$ 1$&$\antifund$&$1$&$ \antifund$ \\
$s_2$&$1$&$\overline\antisym$&$1$ &$1$
\end{tabular}
\ee
\end{center}
and superpotential
\be\label{eq:WK}
W_\text{mag}=\t q a \t q + q M_1 \t q+q s_1 \t p+ \t p a \t p\, s_2+q a^{(N+K-2)/2}M_2 \,.
\ee
Here $N$ is fixed by the gauge group rank of the electric theory, while $K$ is an arbitrary integer with the same parity as $N$. The family of magnetic theories is obtained by varying $K$ over all integers of the prescribed parity.

The salient features in this duality are: \textit{i)} the arbitrarily large magnetic gauge group 
$SU(N+K-1)$, even for fixed electric gauge group (fixed $N$), \textit{ii)} the appearance of a UV $SU(K)$ global symmetry factor that is absent in the electric theory, and \textit{iii)} the presence of a classically irrelevant superpotential interaction. Let us comment more on these points.

One of the lessons of gauge duality has been that the gauge group does not in general define the theory, because the same fixed point can have dual descriptions with different gauge groups. In our dual descriptions we see an extreme version of this, with an infinite set of gauge groups all describing the same infrared dynamics! While in known examples of duality a given electric theory is related to a magnetic theory of fixed gauge group, here we find that an infinite family of theories all describe the same fixed point. In the UV these theories have very different propagating degrees of freedom. For instance, a gauge group $SU(N+K-1)$ has $(N+K-1)^2-1$ gauginos, so increasing $K$ increases the number of degrees of freedom. However, the duality implies that the IR fixed point is actually independent of $K$, so RG effects are responsible for a dramatic reduction in the number of propagating fields, in a way that will be shown explicitly in \S \ref{sec:amax} with $a$-maximization.

On the other hand, global symmetries are believed to be physical and they should match in the electric and magnetic descriptions. However, this need not occur at a perturbative level. Here we will find a novel effect: a global symmetry present classically is removed quantum-mechanically. It will be argued that all the gauge invariants that are charged under the classical $SU(K)$ symmetry of (\ref{table:dualK}) are eliminated from the chiral ring due to nonperturbative superpotentials. Therefore, the $SU(K)$ symmetry does not exist at the quantum level, and the global symmetries of (\ref{table:repeat1}) and (\ref{table:dualK}) will then match in the chiral ring.

Finally, the magnetic duals include superpotential terms that are perturbatively irrelevant, and it is assumed that the model can be UV-completed by a renormalizable theory. A similar situation was encountered in \S \ref{sec:self-dual}. We will find in \S \ref{sec:amax} that the quartic operator is dangerously irrelevant, driving the theory to a fixed point where it becomes marginal. On the other hand, the baryonic deformation will be found to be either dangerously irrelevant or irrelevant, depending on $N$. Notice that the electric theory does not have irrelevant superpotential interactions.

Before proceeding, we point out that $K=1$ recovers the duality of \S \ref{sec:self-dual}. Also, $K=2$ has some special features that are discussed below. The general case corresponds to $K \ge 3$.

\subsection{Duality from product gauge groups}\label{subsec:productK}

Before analyzing the dynamics of (\ref{table:dualK}), we explain some aspects of the correspondence using Seiberg duality. The basic idea is to construct a product gauge group theory that interpolates between the electric and magnetic descriptions in different limits of the ratio of holomorphic dynamical scales. This will be used to deduce that both theories have the same phase structure, because no phase transitions are expected when a holomorphic coupling is varied.

In the context of theories with 2-index representations, this approach was first used by Berkooz~\cite{Berkooz:1995km} for an antisymmetric tensor, and then generalized in~\cite{Pouliot:1995me,Luty:1996cg,Terning:1997jj,Csaki:2004uj}. The process of going from the electric theory to the product gauge group is known as \emph{deconfinement}. Let us summarize this approach in the case relevant for us. Starting from the theory (\ref{table:repeat1}), a new gauge group $Sp(N+K-4)$ is introduced, together with a field $X$, which is a bifundamental of $SU(N)\times Sp(N+K-4)$. Additional fields and interactions are also included, so that the $Sp$ group s-confines and anomalies are canceled.
The antisymmetric tensor $A$ is then identified with a meson of the confining $Sp(N+K-4)$, $A^{\alpha\beta} \sim X^{\alpha \alpha'} X^{\beta \beta'} J_{\alpha' \beta'}$, where $\alpha, \beta$ ($\alpha', \beta'$) are $SU(N)$ ($Sp(N+K-4)$) indices and $J_{\alpha' \beta'}$ 
is the $Sp$ invariant tensor.  

Explicitly, the product gauge group theory has matter content
\begin{center}
\be\label{table:deconf-K}
\begin{tabular}{c|cc|ccc}
&$SU(N)$&$Sp(N+K-4)$&$SU(K)$&$Sp(2N-2)$&$SU(N+3)$  \\
\hline
&&&&\\[-12pt]
$Q$&$\fund$&$1$&$1$&$1$&$\antifund$  \\
$\t Q$&$\antifund$&$1$&$1$&$ \fund$ &$1$ \\
$\t P$&$\antifund$&$1$&$1$&$1$&$1$  \\
$X$&$ \fund$&$\fund$&$ 1$&$1$ &$1$ \\
$U$&$\antifund$&$1$&$\antifund$&$ 1$ &$1$ \\
$V$&$1$&$\fund$&$\fund$&$ 1$ &$1$ \\
$T$&$1$&$1$&$\overline\antisym$&$1$ &$1$
\end{tabular}
\ee
\end{center}
The gauge group is $SU(N) \times Sp(N+K-4)$, while the nonabelian flavor symmetries are $SU(K) \times 
Sp(2N-2) \times SU(N+3)$, where $N$ and $K$ are both either even or odd. Notice the introduction of the 
additional global symmetry $SU(K)$.  The superpotential is taken to be
\be
W=  \t Q XX \t Q+XU V + VVT \,.
\ee
We now study this theory in two different limits (one of which recovers the original electric description), 
depending on which gauge group factor becomes strong first.

\subsubsection{The limit $\Lambda_{Sp(N+K-4)}\gg \Lambda_{SU(N)}$}

If $\Lambda_{Sp(N+K-4)}\gg \Lambda_{SU(N)}$, the strong dynamics of the $Sp$ group dominates first, producing s-confinement~\cite{Intriligator:1995ne}. This yields mesons
\be
\mc M\;:\;(XX)\;,\;(XV)\;,\;(VV)
\ee
and the usual nonperturbative (pfaffian) superpotential. Below the confining scale we obtain
\be
W= \t Q (XX) \t Q+(XV) U + (VV)T + {\rm Pf}\,\mc M\,.
\ee

The fields $(XV)$, $U$, $(VV)$ and $T$ are now massive, and integrating them out we arrive at the electric description (\ref{table:repeat1}). Notice that in this limit all the fields charged under the global $SU(K)$, as well as the $K$ dependence in the gauge group, have disappeared from the low energy theory.

\subsubsection{The limit $\Lambda_{SU(N)}\gg \Lambda_{Sp(N+K-4)}$}

When $ \Lambda_{SU(N)}\gg \Lambda_{Sp(N+K-4)}$ the $SU(N)$ factor should be dualized first. We obtain
\begin{center}
\begin{tabular}{c|cc|ccc}
&$SU(N+K-1)$&$Sp(N+K-4)$&$SU(K)$&$Sp(2N-2)$&$SU(N+3)$  \\
\hline
&&&&\\[-12pt]
$q$&$\fund$&$1$&$1$&$1$&$\fund$  \\
$\t q$&$\antifund$&$1$&$1$&$ \fund$ &$1$ \\
$\t p_1$&$\antifund$&$1$&$1$&$1$&$1$  \\
$x$&$ \fund$&$\fund$&$ 1$&$1$ &$1$ \\
$u$&$\antifund$&$1$&$\fund$&$ 1$ &$1$ \\
$V$&$1$&$\fund$&$\fund$&$ 1$ &$1$ \\
$T$&$1$&$1$&$\overline\antisym$&$1$ &$1$ \\
$(Q \t Q)$&$1$&$1$&$1$&$\fund$ &$\antifund$ \\
$(Q \t P)$&$1$&$1$&$1$&$1$ &$\antifund$ \\
$(Q  U)$&$1$&$1$&$\antifund$&$1$ &$\fund$ \\
$(X \t P)$&$1$&$\fund$&$1$&$1$ &$1$ \\
$(X \t Q)$&$1$&$\fund$&$1$&$\fund$ &$1$ \\
$(X U)$&$1$&$\fund$&$\antifund$&$1$ &$1$
\end{tabular}
\end{center}
The superpotential now reads
\bea
W&=&  (\t Q X) (\t Q X)+(XU)V+VVT+\nonumber\\
&+&q (Q \t Q) \t q+q (Q \t P) \t p_1+q (Q U) u+x (X \t P) \t p_1+x (X \t Q) \t q+x (X U) u\,.
\eea
The terms in the second line arise from Seiberg duality.

Integrating out the heavy fields leaves an s-confining $Sp(N+K-4)$ gauge group, which gives mesons $(xx)$ and $(x(X \t P))$. The confined theory superpotential is
\bea
W&=& \t q (xx)\t q + (xx)uuT+ q (Q \t Q)\t q+q (Q \t P) \t p_1+ q (QU)u +\nonumber\\
&&+ (x(X \t P))\t p_1 + (xx)^{(N+K-2)/2} (x(X \t P))\,,
\eea
where the last term is schematic for the nonperturbative superpotential that contains a Pfaffian of mesons.  
Finally, integrating out $(x(X \t P))$ and $\t p_1$, the low energy theory becomes
\begin{center}
\be\label{table:dualK-temp}
\begin{tabular}{c|c|ccc}
&$SU(N+K-1)$&$SU(K)$&$Sp(2N-2)$&$SU(N+3)$  \\
\hline
&&&\\[-12pt]
$q$&$\fund$&$1$&$1$&$ \fund$  \\
$\t q$&$\antifund$&$1$&$\fund$ &$1$ \\
$u$&$\antifund$&$\fund$&$ 1$ &$1$ \\
$(xx)$&$\antisym$&$ 1$&$ 1$ &$1$ \\
$(Q \t Q)$&$1$&$ 1$&$\fund$&$ \antifund$  \\
$(Q \t P)$&$1$&$ 1$&$1$&$ \antifund$  \\
$(Q U)$&$ 1$&$\antifund$&$1$&$ \antifund$ \\
$T$&$1$&$\overline\antisym$&$1$ &$1$
\end{tabular}
\ee
\end{center}
with
\be\label{eq:WK1b}
W=\t q (xx) \t q +(xx) u u T + q (Q \t Q) \t q+q (Q U) u+q (xx)^{(N+K-2)/2}(Q \t P) \,.
\ee
After a renaming of fields, this coincides with the magnetic description (\ref{table:dualK}). In particular, this procedure explains how the superpotential baryonic deformation arises from the s-confining superpotential.

In summary, starting from the product gauge group theory (\ref{table:deconf-K}), we have shown how to recover the electric and (an infinite family of) magnetic descriptions, by varying the holomorphic ratio $\Lambda_{SU(N)}/\Lambda_{Sp(N+K-4)}$. This result implies that both limits have the same phase structure, and also illustrates how the gauge and global symmetries may be reduced in the IR.

\subsection{Perturbative analysis of the magnetic theory}\label{subsec:pert}

Let us now focus directly on the magnetic description (\ref{table:dualK}). Here we analyze the dynamics at the perturbative level, discussing the classical chiral ring and abelian symmetries. In \S \ref{subsec:nonpert} we include nonperturbative effects and argue that the $SU(K)$ symmetry is eliminated quantum-mechanically.

First we map the anomaly free abelian symmetries of the electric theory (given in (\ref{table:U1-electric})) to the magnetic theory with arbitrary $K$, obtaining
\begin{center}
\begin{tabular}{c|cc}
&$U(1)$&$U(1)_R$  \\
\hline
&&\\[-12pt]
$q$&$-\frac{(K-4) \mc Q_Q+(N+K-4) \mc Q_{\t Q}}{N+K-1}$&$\frac{(N+K-4)(1-R_{\t Q})-(K-4) R_Q}{N+K-1}$  \\
&&\\[-12pt]
$\t q$&$-\frac{(N+3) \mc Q_Q+3 \mc Q_{\t Q}}{N+K-1}$&$\frac{N+K+2-(N+3) R_Q-3 R_{\t Q}}{N+K-1}$ \\
&&\\[-12pt]
$\t p$&$\frac{-(N+3) \mc Q_Q+[ (N-3) +N(N-1)/K] \mc Q_{\t Q}}{N+K-1}$&$R_{\t p}$ \\
&&\\[-12pt]
$a$&$2\frac{ (N+3) \mc Q_Q+3 \mc Q_{\t Q}}{N+K-1}$&$2\frac{(N+3) R_Q+3 R_{\t Q}-3}{N+K-1}$ \\
&&\\[-12pt]
$M_1$&$\mc Q_Q + \mc Q_{\t Q}$&$R_Q+R_{\t Q}$  \\
$M_2$&$-(N+2) \mc Q_Q-2 \mc Q_{\t Q}$&$4-(N+2) R_Q-2 R_{\t Q}$  \\
$s_1$&$\mc Q_Q+\left(1-\frac{N}{K}\right)\mc Q_{\t Q}$&$R_Q+\frac{1}{K}(N+K-2-(N-K) R_{\t Q})$ \\
$s_2$&$-\frac{2 N}{K}\mc Q_{\t Q}$&$\frac{2}{K} (N+K-2-N R_{\t Q})$
\end{tabular}
\end{center}
where
\be
R_{\t p}=\frac{5 -(N-1)(N-2)/K- N-(N+3) R_Q+(N-3+N(N-1)/K) R_{\t Q}}{N+K-1}\,.
\ee

At the perturbative level, the chiral ring of (\ref{table:dualK}) is larger than that of the electric theory, and includes fields that are charged under the $SU(K)$ symmetry. Imposing F-term conditions reduces the gauge invariants in the chiral ring to the singlets
\be
M_1\;,\;M_2\;,\;s_1\;,\;s_2
\ee
and baryons
\be
q^{j+3} a^{(N+K-j-4)/2}\;,\;j\le \;min(N+K-4,\,N)\;,\;j \in 2 \mathbb Z_{\ge 0}\,.
\ee

Using the anomaly free symmetries, the mapping between the two theories for $K \ge 3$ is
\bea\label{eq:match-K}
Q \t Q &\leftrightarrow& M_1\nonumber\\
Q \t P &\leftrightarrow& M_2\nonumber\\
Q^{N-j} A^{j/2} &\leftrightarrow& q^{j+3} a^{(N+K-j-4)/2} \,.
\eea
The two invariants $s_1$ and $s_2$ that are charged under $SU(K)$ do not map to any operator in the electric theory. In \S \ref{subsec:nonpert} it will be argued that these fields are lifted by nonperturbative corrections to the superpotential.

As we mentioned before, the cases $K=1, 2$ are special. For $K=1$ the duality was presented in \S \ref{sec:self-dual}, and there is a complete correspondence between the electric and magnetic chiral ring given in (\ref{eq:match-self}). When $K=2$ ($N$ even) the mapping becomes
\bea\label{eq:match-K2}
Q \t Q &\leftrightarrow& M_1\nonumber\\
Q \t P &\leftrightarrow& M_2\nonumber\\
Q^{N-j} A^{j/2} &\leftrightarrow& q^{j+3} a^{(N-j-2)/2}\;,\;j\le N-2\;,\;j \in 2 \mathbb Z_{\ge 0} \nonumber\\
A^{N/2}& \leftrightarrow& s_2\,.
\eea
The invariant $s_1$ charged under $SU(K=2)$ is not mapped to any electric operator.

\subsection{Nonperturbative effects and truncation of global symmetries}\label{subsec:nonpert}

We reviewed in \S \ref{subsec:consistency} how a field is truncated quantum-mechanically from the chiral ring if its vev leads to a dynamical superpotential that forbids supersymmetric vacua. We now argue that $s_1$ and $s_2$, charged under SU($K$), are removed from the chiral ring by similar effects.

Consider giving $s_1$ a rank $r$ expectation value. A simple way to find the dynamical superpotential is to consider anomalous axial and $R$ symmetries. The relevant fields in the IR dual and their charges under the anomalous symmetries are
\begin{center}
\be\label{table:dualK2}
\begin{tabular}{c|cc}
&$U(1)_A$&$U(1)_R$  \\
\hline
&&\\[-12pt]
$q$&$1$&$-\frac{2}{N+K-1}$  \\
$\t q$&$-1$&$2+\frac{2}{N+K-1}$ \\
$\t p$&$-1$ &$2+\frac{2}{N+K-1}$\\
$a$&$2$ &$-2-\frac{4}{N+K-1}$\\
$\Lambda_L^{N+2K+r-2}$&$N+K-1$ &$0$ \\
$q^{N+3-r} a^{K-2} \t p^{K-r}$&$N+K-1$ &$2-2r$
\end{tabular}
\ee
\end{center}
where $\Lambda_L^{N+2K+r-2}\sim \langle s_1^r \rangle \Lambda^{N+2K-2}$.

For $r>1$, the following superpotential is allowed by all the $U(1)$ symmetries:
\bea
\label{eq:NP2}
W_\text{dyn} =C_{N,K,r} \left(\frac{\Lambda_L^{N+2K+r-2}}{q^{N+3-r} a^{K-2} \t p^{K-r}}\right)^{1/(r-1)}\,,
\eea
where $C_{N,K,r}$ is a nonzero constant (as we argue below).  Eq.~(\ref{eq:NP2}) leads to a runaway with no supersymmetric vacua, so $s_1$ is forced to have rank 1 or less. For $s_1$ of rank 1, similar arguments establish that there is a quantum modified moduli space with supersymmetry breaking. We conclude that $s_1 $ is not part of the chiral ring.

Similarly, when $s_2$ has a rank $2r$ expectation value the following superpotential is consistent with 
all symmetries 
\be
\label{eq:NP3}
W_{np}=C'_{N,K,r} \left(\frac{\Lambda_L^{N+2K+r-2}}{q^{N+3}  a^{K-2-r} \t p^{K-2r}}\right)^{1/(r-1)}\,,
\ee
where $\Lambda_L^{N+2K+r-2}\sim \langle s_2^r \rangle \Lambda^{N+2K-2}$. This leads to a runaway, 
removing $s_2$ from the quantum chiral ring.

The presence of these dynamical effects in the conformal window of the chiral theory is quite intriguing, especially for $s_2$, whose vev does not produce massive quarks (at least at a perturbative level). It would be interesting to check these predictions with an instanton calculation when $r=2$.
However, it is possible to relate these nonperturbative effects in chiral theories (as well as in (\ref{eq:NP1})) to the familiar ADS superpotentials as follows. Consider deconfining the antisymmetric tensor $a$ in the dual theory to obtain a product gauge group theory where the field $a$ of (\ref{table:dualK}) is replaced by an additional gauge group $Sp(N+K-4)$ with matter content
\begin{center}
\be\label{table:deconf}
\begin{tabular}{c|cc}
&$SU(N+K-1)$&$Sp(N+K-4)$  \\
\hline
&&\\[-12pt]
$X'$&$\fund$&$\fund$\\
$U'$&$\antifund$&$1$\\
$V'$&$1$&$\fund$
\end{tabular}
\ee
\end{center}
together with an extra superpotential term 
\be\label{eq:Wdeconf}
W \supset X'U'V'\,.
\ee
The full superpotential of this new theory is (\ref{eq:WK}) plus (\ref{eq:Wdeconf}), replacing $a \to X' X'$. The magnetic theory is recovered in the limit $\Lambda_{Sp(N+K-4)} \gg \Lambda_{SU(N+K-1)}$ when the $Sp(N+K-4)$ factor s-confines.

We first discuss the nonperturbative effects triggered by a nonzero vev of $s_2$.  For this, we take the opposite limit $\Lambda_{Sp(N+K-4)} \ll \Lambda_{SU(N+K-1)}$ and dualize the $SU(N+K-1)$ factor first. The steps are similar to those in \S \ref{subsec:productK}. After dualizing and integrating out massive fields, the matter content is nearly identical to that in (\ref{table:deconf-K}); in this table, 
\{$Q, \t Q, \t P, X, U$\} correspond to the magnetic quarks dual to \{$q, \t q, U', X', \t p$\}, respectively, 
$V$ is the meson $(\t p X')$, and $T$ is $s_2$.  We also have additionally the singlets $M_2$ and the meson 
($q U'$) and two superpotential terms involving these fields, which are unimportant for this discussion.  

At this stage, the $Sp(N+K-4)$ group is s-confining.  Importantly, the meson $(\t p X')$, produced by dualizing the $SU$ factor, is a fundamental flavor of $Sp(N+K-4)$.  If we now turn on a rank $2r$ expectation value for $s_2$, the superpotential term 
\be
W \supset (\t p X') (\t p X') \langle s_2 \rangle
\ee
(see (\ref{eq:WK})) acts as a mass term for $(\t p X')$. Thus below the scale $\langle s_2 \rangle$
the number of fundamentals in the $Sp(N+K-4)$ group is reduced to $N+K-2r$, and for $r>1$ there is a dynamical superpotential~\cite{Intriligator:1995ne}
\be
W_\text{dyn} \propto \left(\frac{\Lambda_{Sp}^{N+K+r-3}}{X^N (\t p X')^{K-2r}} \right)^{1/(r-1)}
\ee
This theory has a runaway instability, with no supersymmetric vacua, truncating $s_2$ from the chiral ring.

A very similar discussion holds when $s_1$ has a nonzero vev.  In this case, we first integrate out the massive $q$ and $\t p$ fields and then dualize the $SU(N+K-1)$.  This also leads to a theory with no supersymmetric vacua. 

Varying now the ratio of scales from $\Lambda_{SU(N+K-1)}/ \Lambda_{Sp(N+K-4)}$ from much greater than one to much smaller than one, we connect these theories without supersymmetric vacua to the magnetic theory  of (\ref{table:dualK}). Since there are no phase transitions under a variation of a holomorphic coupling, we conclude that turning on $s_1$ or $s_2$ leads to a theory (\ref{table:dualK}) without supersymmetric vacua. Therefore $C_{N,K,r}$ and $C'_{N,K,r}$ above are nonzero. Notice that by varying the dynamical scales in the product gauge group theory we have related an Sp instanton calculation to a nonperturbative effect in a chiral theory.

We have thus found that due to dynamical effects, all the gauge invariants that are charged under $SU(K)$ are eliminated from the chiral ring of the magnetic $SU(N+K-1)$ theory. The classical flavor symmetry $SU(K)$ disappears nonperturbatively, and the magnetic global symmetry group that acts on the chiral ring is reduced to that of the original electric theory (\ref{table:repeat1}). This truncation of global symmetries is related to the chiral nature of the models.

\section{Dynamics at the superconformal fixed point}\label{sec:amax}

In \S \ref{sec:K} we argued that the electric theory (which is independent of $K$) and the
magnetic dual theories (for any $K$) flow to the same superconformal fixed point
in the far IR.  The infrared behavior of both theories can be
understood via $a$-maximization~\cite{Intriligator:2003jj}, which will be used in this section to determine the exact dimensions of the gauge invariants in the electric and magnetic theories. The superconformal R-charges will be found to be consistent with the mapping of the chiral rings proposed in (\ref{eq:match-K}). Furthermore, we will show that the $a$-function has the same value in the electric and magnetic theory, giving further
evidence that the two theories are dual.  

In a superconformal theory, the dimension of a gauge invariant operator, $\Delta_\OO$, is proportional to
its superconformal $R$ charge, $R_\OO$; for a spin zero field, the relation is $\Delta_{\mc O}= \frac{3}{2} R_{\mc O}$.
Since there are often many additional $U(1)$ symmetries in the IR, it is not clear
which linear combination of $U(1)$ charges corresponds to the superconformal $R$-charge. This is the case for our theories, since we found in (\ref{table:U1-electric}) that the R-charges are not uniquely determined. 
In~\cite{Intriligator:2003jj} it was shown that the superconformal $R$-charge can be determined
by maximizing the central charge
\be
a = \frac{3}{32} \left[ 3 \Tr R^3 - \Tr R \right]\,.
\ee

The $a$-function is a measure of the number of degrees of freedom
of the theory.
If the electric and magnetic theories are dual, they must describe the same
physics in the far IR, and the propagating degrees of freedom at the fixed point should match. However, the number of degrees of freedom of the magnetic dual theories in the UV depend on $K$ (this can 
be seen from the $a$-function, which explicitly depends on $K$ in the UV when all fields are free and 
have an $R$-charge of 2/3).  This means that the $K$ dependence must be canceled as we flow to the IR fixed point, matching onto the electric theory. The calculation of the value of $a$ at the fixed point will then provide a very nontrivial test on our dual pairs.

We first analyze the electric theory (\ref{table:chiral-self}) in \S \ref{subsec:amax-electric}
and the magnetic dual theory (\ref{table:dualK}) in \S \ref{subsec:amax-magnetic}.

\subsection{$a$-maximization in the electric theory}\label{subsec:amax-electric}

We begin with an analysis of the electric theory. 
To determine the $R$-charges, we use the conditions outlined in footnote \ref{foot:conditions}. 
We self-consistently require each term in the superpotential to be marginal, i.e.~have an $R$-charge of 2, 
and require vanishing of the $U(1)_R SU(N)^2$ anomaly (or equivalently, a vanishing beta function).  
As we found in (\ref{table:U1-electric}), this leaves two unknown charges, which we choose to be
$R_Q$ and $R_{\t Q}$. In terms of these,
\bea \label{eq:RchargesUV}
R_A & = & -2(R_{\t Q} -1) \label{eq:RchargesUV1} \\
R_{\t P} & = & 4-(N+3)R_Q-2 R_{\t Q}\,. \label{eq:RchargesUV2}
\eea

The superconformal R-charges are determined by extremizing the $a$-function~\cite{Intriligator:2003jj}
\bea
a & = & \f{3}{32}\Big(2 (N^2 -1)+N(N+3) f(R_Q) + N f(R_{\t P})  \nonumber \\
& & +N (2N-2) f(R_{\t Q}) + \f{1}{2}N(N-1)f(R_{A}) \Big)\,,  \label{eq:afuncEl}
\eea
where the first term is the contribution from the gauginos, and
\be\label{eq:f(R)}
f(R_\mc{O}) = 3(R_\mc{O}-1)^3 - (R_\mc{O}-1)\,.
\ee

We now use (\ref{eq:RchargesUV1}) and (\ref{eq:RchargesUV2}) to express the 
$a$-function in terms of $R_Q$ and $R_{\t Q}$. Requiring that
\be\label{eq:extr}
\f{\partial a}{\partial R_{Q}} =  0\;,\; \f{\partial a}{\partial R_{\t Q}} = 0\,,
\ee
the solution that maximizes $a$ is
\be
R_{\t Q} = \f{1}{2} \Big(4 - 4 R_Q - N_c R_Q \Big)\,,
\ee
and $R_Q$ can be expressed in terms of $N$ as
\bea
R_Q = \f{12-12N-4N^2+\f{4}{3}\sqrt{N^4+4N^3+5N^2-18N+9}}{12-8N-7N^2-N^3}\,.
\eea
The $R$-charges $R_{\t Q}$, $R_{\t P}$, and $R_A$ can now all be expressed in terms of $N$.  

Next, we have to check that no gauge-invariants in the chiral ring hit the unitarity bound.   For a scalar gauge-invariant operator $\mc{O}$ in the chiral ring, we have to check that $R_\mc{O} > 2/3$.  
A field with $R_\mc{O} < 2/3$ violates unitarity and will instead be interpreted as a free field with  
$R_\mc{O}=2/3$ and decoupled from the superconformal strong dynamics.  Operationally, this has 
implications for the $a$-maximization procedure.  
In particular, there is an accidental $U(1)$ symmetry associated with rotations of the free field $\mc{O}$, which 
needs to be accounted for in the $a$-maximization calculation.  This can be done following~\cite{Kutasov:2003iy,Barnes:2004jj}, modifying the $a$-function as 
\be\label{eq:a-change1}
a  \to  a + \f{3}{32}\,{\rm dim}(\mc{O})\,  f(2/3) - \f{3}{32}\,{\rm dim}(\mc{O})\,  f(R_\mc{O})\,.
\ee

The chiral ring of the theory is given by $Q \t Q$, $Q \t P$, and $Q^k A^{(N-k)/2}$. 
We find that $Q \t P$ violates the unitarity bound for 
\be\label{eq:QPtunit}
N\gtrsim 4.46\,.    
\ee
Above this value, $Q \t P$ is free and is subtracted from the $a$-function,
\bea\label{eq:a-change}
a \to a + \f{{\rm dim}(Q\t P)}{96}\,(2-3R_{Q\t P})^2(5-3R_{Q\t P}),
\eea
where dim$(Q\t P)=N+3$ and $R_{Q\t P}=R_Q + R_{\t P}$.  
The $R$-charges of all the fields can now be determined again from 
(\ref{eq:extr}) (note that (\ref{eq:RchargesUV1}) and (\ref{eq:RchargesUV2}) remain unchanged).  
The explicit solution is somewhat cumbersome, but at large $N$ it simplifies to
\begin{center}
\begin{tabular}{c|cccc}
&$Q$&$\t Q$&$\t P$&$A$\\
\hline
&&&&\\[-12pt]
$R(N \to \infty)$&$ 0.202 $&$0.697$&$  -\f{0.202}{N}$&$0.607$
\end{tabular}
\end{center}
No other fields in the chiral ring hit the unitarity bound as a function of $N$.

The $a$-function at the fixed point can now be expressed in terms of $N$, and at large $N$ it becomes
\be\label{eq:afunc-RM-el}
a \simeq  0.122\,N^2 + \mc O(N)\,.
\ee
Below, we compare the $a$-function of the electric theory to its value in the magnetic dual.

\subsection{$a$-maximization in the magnetic theory for general $K$}\label{subsec:amax-magnetic}

Let us now calculate the $a$-function in the magnetic dual theory for general $N$ and $K$. We will show that the anomalous dimensions and value of the $a$-function are independent of $K$, and agree with those of the electric theory.

Requiring anomaly cancellation and that each term in the superpotential is marginal leaves two unknown $R$-charges. Choosing them to be $R_{M_1}$ and $R_{s_2}$, the rest of the charges read
\bea \label{eq:Rcharges}
R_q & = & -\f{4 + 2(K-4) R_{M_1} - K R_{s_2} }{2(N+K-1)}  \nonumber \\
R_{\t q} & = & -\f{-4(N+K) + 2(N+3) R_{M_1} + K R_{s_2} }{2(N+K-1)}  \nonumber \\
R_a & = & -\f{2(N+K+1) - 2(N+3) R_{M_1} - K R_{s_2} }{N+K-1}  \nonumber \\
R_{s_1} & = & \f{1}{2} (2R_{M_1}+R_{s_2})  \nonumber \\
R_{\t p} & = & -\f{-4(N+K) + 2(N+3) R_{M_1} + (N+2 K-1) R_{s_2} }{2(N+K-1)} \nonumber \\
R_{M_2} & = & \f{1}{2}\big(4+2K+2N-2(N+2) R_{M_1}- K R_{s_2}\big) \,.
\eea

The $a$-function is given explicitly by
\bea
a & = & \f{3}{32}\Big(2 ((N+K-1)^2 -1)+(N+K-1)(N+3)f(R_q) + K(N+K-1)f(R_{\t p})  \nonumber \\
& & \;\;\;\;\;\;\;+(N+K-1)(2N-2)f(R_{\t q}) + \f{1}{2}(N+K-1)(N+K-2)f(R_a)  \nonumber \\
& & \;\;\;\;\;\;\;+K(N+3)f(R_{s_1})+(2N-2)(N+3)f(R_{M_1}) +\nonumber \\
& & \;\;\;\;\;\;\;+\f{1}{2}K(K-1)f(R_{s_2})+(N+3)f(R_{M_2})\Big)\,,
\eea
where $f(R_\OO)$ was defined in (\ref{eq:f(R)}). Using the relations (\ref{eq:Rcharges}) and extremizing, we find the maximum 
\be
R_{s_2} = \f{2}{K(N+2)} \left(-4+2K+(K+4)N +N^2 - 4N(N+1) R_{M_1}\right)\,,
\ee
with
\bea\label{eq:RMgeneral}
R_{M_1} = \f{2\big(-18+15N+6N^2+(N+2) \sqrt{9-18N+5N^2+4N^3+N^4}\big)}{3(-12+8N+7N^2+N^3)}\,.
\eea

We now have to check the unitary bounds for the operators in the chiral ring, given by $M_1$, $M_2$ and 
the baryons $q^{j+3} a^{(N+K-j-4)/2}$. $M_2$ violates the unitarity bound for $N \gtrsim 4.46$. Recalling that $M_2$ is mapped to $Q \t P$, this is precisely the same result we found in (\ref{eq:QPtunit}) for the electric theory, which is quite encouraging for our duality. 

Above this value of $N$, $M_2$ becomes free and decouples from the interacting sector. This implies that the last term of the superpotential (\ref{eq:WK}) becomes irrelevant in this range. We conclude that this baryonic operator changes from dangerously irrelevant to irrelevant, as $N$ is increased to $N \gtrsim 4.46$. Thus, above this value the $a$-maximization computation has to be corrected in two ways: the last term in the superpotential has to be ignored (it becomes irrelevant), and the contribution of $M_2$ has to be subtracted as we explained before. The large $N$ values of the corrected $R$-charges are
\begin{center}
\begin{tabular}{c|cccccccc}
&$q$&$\t q$&$\t p$&$a$&$M_1$&$M_2$&$s_1$&$s_2$\\
\hline
&&&&&&&\\[-12pt]
$R(N \to \infty)$&$0.303$&$0.798$&$ -0.303 \,\f{N}{K}$&$0.403$&$0.898$&$\frac{2}{3}$&$0.303\,\f{ N}{K}$&$0.607 \,\f{N}{K}$
\end{tabular}
\end{center}

Expressing the $a$-function in terms of $N$ and $K$, remarkably the $K$ dependence
cancels completely, and the expression agrees exactly with the expression of the electric theory for all 
$N$ and $K$! In summary, we have shown that the $R$-charges and values of the $a$-functions of the electric and magnetic theories agree, providing highly nontrivial evidence for the validity of the duality.

\section{The phase structure for $F<N+3$ flavors}\label{sec:mass}

Having understood the superconformal phase that arises when $F=N+3$, in this section we analyze the phase structure when the number of flavors is reduced to $F \le N+2$. Mass terms can be added to decouple $\t P$ and/or some number of $\t Q$ flavors. These cases are studied separately, as they lead to different infrared dynamics. We will find a rich phase structure, including a conformal fixed point, confinement with and without chiral symmetry breaking, and finally a nonperturbative instability for $F \le N$.

\subsection{$F=N+2$ flavors from integrating out $\t P$ and one $Q$}\label{subsec:tP}

Let us first study the mass flow upon deforming the theory with a mass for a single flavor consisting of the 
fundamentals $\t P$ and $Q_{N+3}$,
\be\label{eq:elecm}
\delta W_\text{elec} = m\, Q_{N+3} \t P\,,
\ee
where $Q_{N+3}$ is the $(N+3)$th flavor of $Q$. Below the scale $m$, the electric theory becomes
\begin{center}
\be\label{table:onelessflavor}
\begin{tabular}{c|c|cc}
&$SU(N)$&$Sp(2N-2)$&$SU(N+2)$  \\
\hline
&&&\\[-12pt]
$Q$&$\fund$&$1$&$\antifund$  \\
$\t Q$&$\antifund$&$ \fund$ &$1$ \\
$A$&$\antisym$&$1$&$1$  \\
\end{tabular}
\ee
\end{center}
where $Q$ has been reduced to $N+2$ flavors. The superpotential is still $W_\text{el}=  \t Q A \t Q$.

The electric theory below the scale $m$ coincides with the electric theory of a new confining duality recently proposed in~\cite{Spiridonov:2009za}.\footnote{We take care to observe that the proposed duality of \cite{Spiridonov:2009za} does not explicitly include the electric superpotential $\t Q A \t Q$. However, such a superpotential is consistent with the explicit $Sp(2N-2) \times SU(N+2)$ global symmetry, so its inclusion does not change the superconformal index \cite{Romelsberger:2007ec}. Moreover, the theory {\it without} such a superpotential would enjoy an  $SU(2N-2) \times SU(N+2)$ global symmetry, and hence a different index and dual description from that presented in  \cite{Romelsberger:2007ec}. Thus we posit that the new confining duality of \cite{Romelsberger:2007ec} implicitly includes the electric superpotential deformation $\t Q A \t Q$. }  We will find that the mass flow in the magnetic theory coincides with the magnetic theory in~\cite{Spiridonov:2009za}, providing a further non-trivial test of our proposed duality.

\subsubsection{Magnetic description}

We begin with the magnetic theory in \S \ref{sec:self-dual} (i.e.~$K=1$ in the magnetic theory (\ref{table:dualK})).  
The electric deformation (\ref{eq:elecm}) corresponds in the magnetic theory to a deformation
\be
\delta W_\text{mag} = m \Lambda (M_2)_{N+3}
\ee
where $(M_2)_{N+3}$ refers to the $(N+3)$th flavor of the meson $M_2 \sim (Q \tilde P)$, and we are taking $\Lambda_\text{elec} = \Lambda_\text{mag} = \Lambda$ for simplicity. The equations of motion for $M_2$ then set
\be
\langle q_{N+3} a^{(N-1)/2} \rangle = - m \Lambda.
\ee
The combination of $F$- and $D$-terms higgses the magnetic gauge group to $Sp(N-1)$. Expanding around this vacuum, various fields obtain masses of order $(m \Lambda^{\frac{N-1}{2}})^{\frac{2}{N+1}}$ via (\ref{eq:oddWmag}); the only remaining light degrees of freedom are 
\begin{center}
\be\label{table:mag-oneless}
\begin{tabular}{c|c|cc}
&$Sp(N-1)$&$Sp(2N-2)$&$SU(N+2)$  \\
\hline
&&&\\[-12pt]
$q$&$\fund$&$1$&$ \fund$  \\
$\t p$&$\fund$&$ 1$ &$1$ \\
$M_1$&$1$&$\fund$&$ \antifund$ \\
$s$&$1$&$1$&$ \antifund$\end{tabular}
\ee
\end{center}
where $M_2, \tilde q,$ and some components of the remaining fields have been rendered massive, reducing the second global symmetry to $SU(N+2)$. The superpotential is (after a field redefinition to absorb factors of $\langle a \rangle$)
\be
W_\text{mag}= q M_1 M_1 q + q s \t p.  
\ee 
This $Sp(N-1)$ theory then $s$-confines, giving mass to $(q \tilde p), s$ and leaving 
 \begin{center}
\be\label{table:yay}
\begin{tabular}{c|cc}
&$Sp(2N-2)$&$SU(N+2)$  \\
\hline
&&\\[-12pt]
$(qq)$&$1$&$ \antisym$  \\
$M_1$&$\fund$&$ \antifund$ \\
\end{tabular}
\ee
\end{center}
with superpotential
\be\label{eq:Wyay}
W_\text{mag} = M_1 M_1 (qq)
\ee
which coincides with the new confining dual of~\cite{Spiridonov:2009za}. Here $Sp(2N-2)\times SU(N+2)$ are global symmetries. The magnetic description implies that the IR phase corresponds to confinement without chiral symmetry breaking.

At the origin of moduli space, the mapping of chiral rings is
\be
Q \t Q \leftrightarrow M_1\;,\; Q^N \leftrightarrow (qq)\,.
\ee
The remaining electric baryons $Q^{N-j} A^{j/2}$ are not part of the chiral ring of the theory at the IR fixed point.  These are removed by nonperturbative effects, though a detailed study will not be presented here. Instead, these effects will be illustrated in the case $N=3$, which we will discuss now.  

It is useful to consider the case $N=3$ in more detail, because the electric theory becomes SQCD with $SU(3)$ gauge group, 5 flavors (the antisymmetric is equivalent to an antifundamental), and a baryonic deformation.
We can then independently derive a magnetic description using Seiberg duality, and compare with our proposed duality. The Seiberg dual of (\ref{table:onelessflavor}) is $SU(2)$ with five flavors and a superpotential
\be
W_\text{mag}= q (Q \t Q) \t q + q (Q A) a + \t q^2\,,
\ee
where $q$ is the magnetic quark dual to $Q$, and both $A$ and $a$ denote antifundamentals (in the electric and magnetic theory, respectively). The last term in this superpotential is the magnetic baryon $\t q^2$ dual to $\t Q A \t Q$. This is now a mass term; integrating out $\t q$ we are left with $SU(2)$ with 3 flavors and a superpotential $W= q (Q \t Q) q (Q \t Q)+ q (QA) a$, which agrees with (\ref{table:mag-oneless}) at $N=3$. This theory s-confines, giving masses to $(qa)$ and $(QA)$, thus reproducing our dual (\ref{eq:Wyay}).

We have focused on the case of $K=1$ for clarity, but it is useful to consider the mass flow in the magnetic theory for general $K$. The flow proceeds in an entirely analogous fashion, with the gauge group higgsed to an s-confining $Sp(N+K-2)$ theory. After s-confinement, the superpotential lifts the fields $(q \t p), (\t p \t p), s_1, s_2$, eliminating all fields charged under $SU(K)$ and reducing the IR matter entirely to (\ref{table:yay}). Thus we see that the infinite family of magnetic theories parameterized by $K$ flows to a single theory under mass deformation.

The proposed duality between (\ref{table:onelessflavor}) and (\ref{table:yay}) may also be checked using a product gauge group theory, and considering different limits of the holomorphic scales, along the lines of \S \ref{subsec:productK}. This calculation is discussed in the Appendix.

\subsubsection{Phase structure}

The magnetic description (\ref{table:yay}) is a weakly coupled theory of singlets, without gauge interactions. Thus the theory is in a confining phase, and it does not break chiral symmetry.

The IR dynamics can be studied directly using $a$-maximization.  This allows us to determine the 
$R$-charges as described in \S \ref{sec:amax}.
In the electric theory (\ref{table:onelessflavor}), assuming vanishing $\beta$-function and a marginal 
superpotential, we find 
\bea
R_Q = \f{2}{3N}\,, \hskip 1cm 
R_{\t Q} = \f{2}{3} - \f{3}{N}\,, \hskip 1cm 
R_A = \f{2}{3} + \f{4}{3N}\,.
\eea
The $R$-charges of the gauge invariants are therefore 
\bea
R_{Q \t Q} = \f{2}{3}\,, \hskip 1cm R_{Q^{N-j}A^{j/2}} = \f{2+j}{3} \hskip 0.5cm (j\in 0, 2, \ldots, N).  
\eea
This shows that the fields $Q\t Q$ and $Q^N$ are at the unitarity bound and are free fields, consistent with a confined description in terms of weakly coupled gauge singlets. The total $a$-function is $a = (5N^2+7N-6)/96$.

Similarly, in the magnetic theory (\ref{table:yay}) we obtain
\be
R_{M_1} = \f{2}{3}\,,\hskip 1cm
R_{(qq)} = \f{2}{3}\,.
\ee
The fields $M_1$ and $(qq)$ are free, and consistent with the mapping to $Q \t Q$ and $Q^N$, respectively.  The $a$-functions of both the 
electric and magnetic theories again agree exactly.

It is interesting to contrast the resulting phase diagram of the chiral theory with that of SQCD. In the case of SQCD, the self-dual point in the interacting window corresponds to $N_f = 2 N_c$, below which there is an IR-free phase for $N_c +2 \le N_f \leq \frac{3}{2} N_c$, followed by confinement without chiral symmetry breaking (s-confinement) at $N_f=N_c+1$, and confinement with chiral symmetry breaking if $N_f = N_c$. In the case of the chiral theories studied here -- for any value of $K$ -- integrating out a single flavor changes the phase from conformal directly to s-confining. A similar behavior was found in other chiral examples in~\cite{Terning:1997jj,Karch:1997jp}. Below we will obtain a transition from conformal to confining with chiral symmetry breaking (as is expected in QCD).

\subsubsection{The superconformal index}

Further evidence for duality may be obtained by studying the superconformal indices \cite{Romelsberger:2005eg} of our proposed dual theories. The equality of superconformal indices in theories related by Seiberg duality was conjectured in \cite{Romelsberger:2007ec}, and impressive evidence has been found by comparing superconformal indices of known Seiberg duals \cite{Spiridonov:2009za}. While it is beyond the scope of this work to carry out an explicit computation of the superconformal index for the infinite family of dual theories under consideration, there is considerable evidence to support our conjecture that the superconformal indices of electric and magnetic theories agree. 

As we have just seen, upon adding a mass deformation for one flavor our electric and magnetic theories with $F=N+3$ flow to the electric and magnetic theories of \cite{Spiridonov:2009za} with $F=N+2$, for which the superconformal indices were found explicitly to agree. While this alone is not sufficient to demonstrate agreement between the superconformal indices of the undeformed theories, it is quite suggestive.

Perhaps more compellingly, we found in \S\ref{subsec:productK} that the electric and magnetic theories with $F=N+3$ may be related via a product gauge group in the UV. It is expected that the superconformal indices of such theories will agree~\cite{Spiridonov:2009za}. In this case, the equality of superconformal indices follows from Bailey-type chains of symmetry transformations~\cite{Bailey} relating the corresponding elliptic hypergeometric integrals. Indeed, further support for this conjecture regarding the superconformal indices of theories related by deconfinement may be found in the Appendix. There we connect the proposed ($F=N+2$) dual theories of~\cite{Spiridonov:2009za}  via a product theory in the UV, consistent with the agreement of their
 superconformal indices.  

\subsection{$F=N+2$ flavors from integrating out one $Q$ and one $\t Q$ }\label{subsec:tQ}

A different theory with $F=N+2$ is obtained by giving a mass to one $Q$ and one $\t Q$,
\be
\delta W_\text{el} =m Q^\alpha_{N+3} \t Q_\alpha^{2N-2}\,.
\ee
Below the scale $m$, the F-term for $\t Q_\alpha^{2N-2}$ implies that $\t Q^\alpha_{2N-3}$ decouples from the superpotential.
The theory then acquires an accidental $SU(2)$ global symmetry that acts on $\t P$ and $\t Q_{2N-3}$. Combining these fields into a doublet (denoted again by $\t P$), the matter content is
\begin{center}
\be\label{table:chiral-selfNplus2}
\begin{tabular}{c|c|ccc}
&$SU(N)$&$SU(2)$&$Sp(2N-4)$&$SU(N+2)$  \\
\hline
&&&&\\[-12pt]
$Q$&$\fund$&$1$&$1$&$\antifund$  \\
$\t Q$&$\antifund$&$1$&$ \fund$ &$1$ \\
$\t P$&$\antifund$&$\fund$&$1$&$1$  \\
$A$&$\antisym$&$1$&$1$&$1$  \\
\end{tabular}
\ee
\end{center}
with superpotential $W_\text{el}=  \t Q A \t Q$. Notice that now $\t Q$ refers only to the remaining $2N-4$ light flavors.

\subsubsection{Magnetic description}

In the magnetic dual (\ref{table:dualK}), the electric mass term maps to a linear deformation
\be
\delta W_\text{mag} = m \Lambda (M_1)^{2N-2}_{\phantom{c}N+3}
\ee
that induces an expectation value $( q \t q)^{2N-2}_{\phantom{c}N+3}=-m \Lambda$.
This breaks the gauge group to $SU(N+K-2)$, and the Higgsing may be chosen along 
\be
q^{N+K-1}_{\phantom{c}N+3}= \t q^{2N-2}_{\phantom{c}N+K-1}=\sqrt{-m \Lambda}\,.
\ee 

Decomposing the fields in representations of this group and integrating out massive matter, the low energy theory becomes
\begin{center}
\be\label{table:dualKNplus2}
\begin{tabular}{c|c|cccc}
&$SU(N+K-2)$&$SU(K)$&$SU(2)$&$Sp(2N-4)$&$SU(N+2)$  \\
\hline
&&&&\\[-12pt]
$q$&$\fund$&$1$&$1$&$1$&$ \fund$  \\
$\t q$&$\antifund$&$1$&$1$&$\fund$ &$1$ \\
$\t p$&$\antifund$&$\fund$&$1$&$ 1$ &$1$ \\
$a$&$\antisym$&$ 1$&$1$&$ 1$ &$1$ \\
$M_1$&$1$&$ 1$&$1$&$\fund$&$ \antifund$  \\
$N_1$&$1$&$ 1$&$\fund$&$1$&$ \antifund$  \\
$N_2$&$1$&$ 1$&$1$&$1$&$ 1$  \\
$s_1$&$ 1$&$\antifund$&$1$&$1$&$ \antifund$ \\
$s_2$&$1$&$\overline\antisym$&$1$&$1$ &$1$
\end{tabular}
\ee
\end{center}
where $(N_1)^a_{\phantom{c}i}$ combines $(M_2)_i$ and $(M_1)^{2N-3}_{\phantom{c}i}$ into an $SU(2)$ doublet ($a=1,2$ and $i=1,\ldots, N+2$), while $N_2 \propto (M_2)_{N+3}$. The superpotential now reads
\be\label{eq:WmagN2tQ}
W_\text{mag}= \t q a \t q+ q M_1 \t q+ q s_1 \t p+\t p a \t p s_2 + a^{(N+K-2)/2}N_2+ q^2 a^{(N+K-4)/2} N_1^2\,.
\ee 
The last two terms come from the baryonic deformation $q a^{(N+K-2)/2}M_2$ in (\ref{eq:WK}), after integrating out the massive fields $q^\alpha_{N+3}$ and $a_{N+K-1,\alpha}$. The mapping of the chiral rings can be worked out starting from (\ref{eq:match-K}) and integrating out massive fields; in particular we note that
\be\label{eq:map-2less}
Q \t Q \leftrightarrow M_1\;,\;Q \t P \leftrightarrow N_1\;,\;\t P A \t P \leftrightarrow N_2\,.
\ee

Additional evidence for this duality is obtained, as before, by constructing a product gauge group theory that interpolates between the electric and magnetic descriptions by varying a ratio of holomorphic parameters (see the Appendix). Note that the magnetic theory depends on the arbitrary integer $K$, as in the duals found in \S \ref{sec:K}. The $a$-maximization results below show that the dynamics at the fixed point is actually independent of $K$. 

An interesting case arises for $K=2$, because the electric and magnetic gauge groups coincide, giving a self-dual description. This is valid for even $N$, and complements the self-dual theory found in \S \ref{sec:self-dual} for odd $N$ and $F= N+3$ flavors.

\subsubsection{Phase structure}

Let us now discuss the long-distance dynamics using $a$-maximization.  The superconformal $R$-charges in the electric theory (\ref{table:chiral-selfNplus2}) are
\be
R_Q = \f{2}{N+4}\,, \hskip 1cm 
R_{\t Q} = \f{2}{3}\,, \hskip 1cm 
R_{\t P} = \f{2}{N+4}\,, \hskip 1cm 
R_A = \f{2}{3}\,.
\ee 
We see that $Q\t P$ hits the unitarity bound for all $N\ge 2$.  For larger $N$, the corrected procedure yields
\be
R_Q=\f{-3+\sqrt{8N+9}}{3N}\;,\;R_{\t Q}=R_A=\f{2}{3}\;,\;R_{\t P}=\f{-3(3N+2) + (N+2)\sqrt{8N+9}}{6N}\,.
\ee
The next field to hit the unitarity bound for $N\gtrsim (5+3\sqrt{5})/2 \simeq 5.85$ is $\t P A \t P$.  
Re-doing the $a$-maximization procedure with $\t P A \t P$ removed, we can again find the $R$-charges, 
which are too cumbersome to display here.  No other gauge invariants hit the unitarity bound.  

Now consider the magnetic dual theory (\ref{table:dualKNplus2}).  We find that 
\be
R_{N_1}=\f{4}{N+4}
\ee 
so $N_1$ 
hits the unitarity bound for $N\ge 2$.  This is consistent with the above result for $Q \t P$ and the map (\ref{eq:map-2less}).
It also means that the last term in the superpotential is irrelevant for $N\ge 2$.  Ignoring this last term 
and removing $N_1$ from the $a$-maximization procedure, we can determine the $R$-charges for 
$N\ge 2$.  In particular,
\be
R_{N_2} = \f{11 N + 6-(N+2)\sqrt{8N+9}}{3N}
\ee
so $N_2$ hits the 
unitarity bound for $R_{N_2}=(5+3\sqrt{5})/2\simeq 5.85$, just like the gauge invariant $\t P A \t P$ in the 
electric theory, to which $N_2$ gets mapped.  
The second-to-last term in (\ref{eq:WmagN2tQ}) thus becomes irrelevant for $N\gtrsim 5.85$, although it 
is interesting to note that in the range $2 \leq N \leq 5$ it is dangerously irrelevant.  

The $a$-maximization procedure can be continued for larger $N$, and we find no other fields hitting the unitarity bound.  Although we will not show the remaining $R$-charges and values of the $a$-functions, we have checked that the predictions from the electric and magnetic theories agree at the superconformal fixed point.

These results show that the theory is in a conformal phase.

\subsection{The theory with $F<N+2$ flavors}

Integrating out one more flavor, we flow to a theory with $F=N+1$ and matter content\footnote{We have performed an appropriate field renaming that depends on whether we integrate out two $Q$'s and 
either two $\t Q$'s or one $\t Q$ and one $\t P$.}
\begin{center}
\be\label{table:twolessflavors}
\begin{tabular}{c|c|cc}
&$SU(N)$&$Sp(2N-4)$&$SU(N+1)$  \\
\hline
&&&\\[-12pt]
$Q$&$\fund$&$1$&$\antifund$  \\
$\t Q$&$\antifund$&$ \fund$ &$1$ \\
$\t P$&$\antifund$&$ 1$ &$1$ \\
$A$&$\antisym$&$1$&$1$  \\
\end{tabular}
\ee
\end{center}
with $W_\text{el}=  \t Q A \t Q$.

The magnetic dual can be obtained in three different ways, by turning on mass deformations in (\ref{table:yay}), (\ref{table:dualKNplus2}), or the product gauge group described in the Appendix. All these flows consistently lead to the same low energy magnetic dual, which is a theory with no gauge group and weakly coupled fields. For instance, starting from (\ref{table:yay}), the electric mass term corresponds to $\delta W_\text{mag}=m\Lambda (M_1)_{2N-2}^{N+2}$, which induces an expectation value
\be\label{eq:vev2less}
\langle (qq)_{N+2,N+1}(M_1)_{2N-3}^{N+1}\rangle = -m \Lambda\,.
\ee
Integrating out massive matter, the desired magnetic dual is
\begin{center}
\be\label{table:yay2}
\begin{tabular}{c|cc}
&$Sp(2N-4)$&$SU(N)$  \\
\hline
&&\\[-12pt]
$(qq)$&$1$&$ \antisym$  \\
$M_1$&$\fund$&$ \antifund$ \\
$(qq)_{N+2,i}$&$1$&$ \fund$ \\
$(M_1)_{2N-3}^i$&$\fund$&$ \antifund$ \\
$S$&$1$&$1$
\end{tabular}
\ee
\end{center}
with superpotential
\be\label{eq:Wyay2}
W_\text{mag} = M_1 M_1 (qq)\,.
\ee
The singlet $S$ corresponds to the complex modulus that parametrizes (\ref{eq:vev2less}).

This shows that when $F=N+1$ the theory is in a confining phase, with chiral symmetry breaking:  the global chiral symmetry is broken from $Sp(2N-4)\times SU(N+1)$ to $Sp(2N-4)\times SU(N)$.  
As before, this may also be checked using $a$-maximization. Notice that we started from an s-confining theory and ended up in a confining phase with chiral symmetry breaking.  Starting instead from the theory (\ref{table:chiral-selfNplus2}) and turning on a mass term, we obtain a transition from conformal to confining with chiral symmetry breaking. These types of transitions are very interesting because of their potential connection to nonperturbative effects in QCD.

It is interesting to contrast the chiral symmetry breaking above with the case $N_f=N_c$ in $SU(N_c)$ SQCD, where 
a quantum modified moduli space breaks the chiral symmetry.  A similar effect must occur in the electric theory here (\ref{table:twolessflavors}), although
we have not calculated this directly. The chiral symmetry breaking can be seen explicitly in the magnetic dual (\ref{table:yay2}).  Moreover, using the product gauge group in Appendix \ref{sec:prodflow} and setting $F=N+1$, the chiral symmetry breaking is seen to originate from a quantum constraint generated by the Sp dynamics.  

Finally, for $F \le N$, we find that the theory develops a runaway instability, caused by nonperturbative effects. The considerations here are similar to those in \S \ref{subsec:nonpert}. Using holomorphy and symmetries, it can be shown that the electric theory with $F \le N$ allows for a nonperturbative superpotential with runaway behavior. In terms of the product gauge group theory of the Appendix (now for $F\le N$), these nonperturbative effects can be reproduced from an Sp ADS superpotential. This ends our analysis of the phase structure of the chiral theories with $F \le N+3$.

\section{Generalizations and chiral/nonchiral dualities}\label{sec:Kpr}

In this section, we consider generalizations and applications of the previously discussed dualities. 
In \S \ref{subsec:general}, we present an infinite family of electric theories that is dual to an 
infinite family of magnetic theories. These theories have different gauge groups and 
perturbative flavor symmetries, and all flow to the same IR fixed point. Using these 
results, in \S \ref{subsec:nonchiral} we find a new chiral/nonchiral duality, relating a theory with an 
antisymmetric and (anti)fundamentals to another theory with just (anti)fundamentals and singlets.

\subsection{An infinite family of duals}\label{subsec:general}

Having found that a single electric theory can be dual to an infinite family of magnetic descriptions, a natural question is whether this can be extended to a duality between infinite families of both electric and magnetic theories. We now exhibit this phenomenon in a class of theories similar to (\ref{table:repeat1}), albeit with the addition of extra singlets and interactions.

The electric theory is given by
\begin{center}
\be\label{table:Kel}
\begin{tabular}{c|c|ccc}
&$SU(N+K-1)$&$SU(K)$&$Sp(2N-2)$&$SU(N+3)$  \\
\hline
&&&&\\[-12pt]
$Q$&$\fund$&$1$&$1$&$\antifund$  \\
$\t Q$&$\antifund$&$1$&$ \fund$ &$1$ \\
$\t P$&$\antifund$&$\fund$&$1$&$1$  \\
$A$&$\antisym$&$1$&$1$&$1$  \\
$S_1$&$1$&$\antifund$&$1$&$\fund$ \\
$S_2$&$1$&$\overline \antisym$&$1$&$1$
\end{tabular}
\ee
\end{center}
with a superpotential
\be\label{eq:WKel}
W_\text{el}= \t Q A \t Q+Q \t P S_1+\t P A \t P S_2\,.
\ee
The gauge group is $SU(N+K-1)$ and the rest of the groups are flavor symmetries. The theory has a nonrenormalizable interaction that is actually dangerously irrelevant, as in (\ref{eq:WK}).

On the other hand, consider a magnetic theory with matter
\begin{center}
\be\label{table:Kmag}
\begin{tabular}{c|c|ccc}
&$SU(N+K'-1)$&$SU(K')$&$Sp(2N-2)$&$SU(N+3)$  \\
\hline
&&&\\[-12pt]
$q$&$\fund$&$1$&$1$&$ \fund$  \\
$\t q$&$\antifund$&$1$&$\fund$ &$1$ \\
$\t p$&$\antifund$&$\fund$&$ 1$ &$1$ \\
$a$&$\antisym$&$ 1$&$ 1$ &$1$ \\
$M$&$1$&$ 1$&$\fund$&$ \antifund$  \\
$s_1$&$ 1$&$\antifund$&$1$&$ \antifund$ \\
$s_2$&$1$&$\overline\antisym$&$1$ &$1$
\end{tabular}
\ee
\end{center}
and superpotential
\be\label{eq:WKmag}
W_\text{mag}=\t q a \t q + q M \t q+q s_1 \t p+ \t p a \t p\, s_2 \,.
\ee
Here $K'$ is any integer such that $N+K+K'$ is odd. We propose that the infinite set of electric theories with fixed $N$ and arbitrary $K$ is dual to the family of magnetic theories with the same $N$ and arbitrary $K'$ (of the allowed parity).

\subsubsection{Tests of the duality}

As before, we may perform various different tests on the conjectured duality. Here we summarize briefly some of these.

First, the map of the electric and magnetic chiral rings for $K, K' \ge 3$ is
\bea\label{eq:ringKKpr}
Q^{N+K-1-j} A^{j/2}&\leftrightarrow& q^{j-K+4} a^{(N+K+K'-5-j)/2}\,,\;\;\;\;\;\;K-4\le j \le N+K-1\nonumber\\
Q \t Q &\leftrightarrow& M \,.
\eea
The rest of the mesons of the electric theory are lifted by the classical superpotential, and all of the mesons in the magnetic theory are lifted. Nonperturbative superpotentials imply that $S_i$ and $s_i$ are not part of the chiral ring. Therefore, similarly to what we found before, matter charged under the global symmetry factors $SU(K)$ and $SU(K')$ are dynamically eliminated from the chiral ring. Additional evidence follows from the approach of \S \ref{sec:K}, which can be used to construct a product gauge group theory that interpolates between the electric and magnetic description as the ratio of holomorphic scales is varied.

The values $K,\,K' = 1,\,2$ are special cases, because some of the $S_i$ or $s_i$ become part of the chiral ring. For instance, for $K=1$ the correspondence is
\bea\label{eq:matchspecial}
Q^{N-j} A^{j/2} &\leftrightarrow& q^{j+3} a^{(N+K'-j-4)/2}\nonumber\\
S_1 &\leftrightarrow& q a^{(N+K'-1)/2}\nonumber\\
Q \t Q &\leftrightarrow& M\,.
\eea

We have also found agreement between the electric and magnetic predictions using $a$-maximization. The superconformal $R$-charges are consistent with (\ref{eq:ringKKpr}). Moreover, we self-consistently find that the perturbatively irrelevant quartic superpotential terms in the electric and magnetic theory are marginal at the strongly coupled fixed point. All of the fields are above the unitarity bound for any allowed $N$, $K$, and $K'$. 

Furthermore, the $a$-function in the electric and magnetic theories are the same and 
independent of both $K$ and $K'$, which provides additional evidence 
that these theories all flow to the same IR fixed point for arbitrary allowed $K$ and $K'$.  
This reveals interesting strong coupling effects, since both families of theories can independently have an arbitrary amount of 
matter in the UV, and yet the number of degrees of freedom at the fixed point, as measured by the $a$-function, is the same.  

\subsubsection{Adding baryonic deformations}

The duality can be extended by adding deformations to the electric and magnetic descriptions. To make a closer contact with the results of \S \ref{sec:K}, we can deform (\ref{table:Kel}) by an additional operator
\be\label{eq:Weldef}
W_\text{el} \supset M_2\,Q^{N+2} A^{(K-3)/2}\,,
\ee
where $M_2$ is a new singlet under the gauge group and transforms as an antifundamental of the flavor $SU(N+3)$.

Using the mapping of baryons (\ref{eq:ringKKpr}), we find that the dual magnetic theory (\ref{table:Kmag}) contains the additional singlets $M_2$ and is deformed by
\be
W_\text{mag}\supset M_2\,q a^{(N+K'-2)/2}\,.
\ee
This theory is the same as the magnetic dual found in \S \ref{sec:K} (renaming $K \leftrightarrow K'$), 
while the electric description in that duality has now been generalized to (\ref{table:Kel}) (together with the 
extra $M_2$ and the deformation (\ref{eq:Weldef})).

The $a$-function for the magnetic dual theory was already calculated in \S \ref{subsec:amax-magnetic}.  
We checked that it agrees with the electric theory with the baryonic deformation.  
As in the magnetic theory, the baryonic term in the superpotential in the electric theory is dangerously 
irrelevant for $N\lesssim 4.46$ and becomes marginal for $N\gtrsim 4.46$. 
It is interesting that these statements hold for arbitrarily allowed $K$ and $K'$ -- in particular, for $N<4.46$, 
the baryonic term can classically look highly irrelevant, yet still be marginal. 

\subsection{Chiral/nonchiral dualities}\label{subsec:nonchiral}

We end our analysis by pointing out that the case $N=3$, $K'=1$ in the above duality leads to an interesting chiral/nonchiral duality. The reason is that the magnetic gauge group then becomes $SU(3)$, and the antisymmetric is simply an antifundamental. Let us briefly summarize this case.

The electric theory is a chiral model with content (after renaming $N_c=K+2$)
\begin{center}
\be\label{table:Kel3}
\begin{tabular}{c|c|ccc}
&$SU(N_c)$&$SU(N_c-2)$&$Sp(4)$&$SU(6)$  \\
\hline
&&&&\\[-12pt]
$Q$&$\fund$&$1$&$1$&$\antifund$  \\
$\t Q$&$\antifund$&$1$&$ \fund$ &$1$ \\
$\t P$&$\antifund$&$\fund$&$1$&$1$  \\
$A$&$\antisym$&$1$&$1$&$1$  \\
$S_1$&$1$&$\antifund$&$1$&$\fund$ \\
$S_2$&$1$&$\overline \antisym$&$1$&$1$ \\
\end{tabular}
\ee
\end{center}
with
\be\label{eq:WKel2}
W_\text{el}= \t Q A \t Q+Q \t P S_1+\t P A \t P S_2\,.
\ee

The magnetic dual has matter content
\begin{center}
\be\label{table:Kmag4}
\begin{tabular}{c|c|cc}
&$SU(3)$&$Sp(4)$&$SU(6)$  \\
\hline
&&&\\[-12pt]
$q$&$\fund$&$1$&$ \antifund$  \\
$\t q$&$\antifund$&$\fund$ &$1$ \\
$\t p$&$\antifund$&$ 1$ &$1$ \\
$\t r$&$\antifund$&$ 1$ &$1$ \\
$s$&$ 1$&$1$&$ \fund$
\end{tabular}
\ee
\end{center}
and interactions
\be\label{eq:WKmag4}
W_\text{mag}=\t q \t p \t q + q \t r s\,.
\ee
This description may be obtained from (\ref{table:Kmag}) after setting $N=3$, $K'=1$, 
applying Seiberg duality once, and integrating out the massive matter 
(we do not show the intermediate steps).  The extra antifundamental $\t r$ arises from 
the antisymmetric, while $s$ is the meson $(qa)$ and the other fields are dual quarks that 
have been re-labelled and given the same notation as the original magnetic quarks in (\ref{table:Kmag}).
The superpotential term $\t q a \t q$ of the magnetic theory in (\ref{eq:WKmag}) is now a baryon 
that gets mapped to the first term in the Seiberg dual (\ref{eq:WKmag4}).

This duality relates a theory with an antisymmetric tensor and some (anti) fundamentals, to another theory with only (anti)fundamentals. As such, it may have applications to particle physics models based on $SU(5)$ GUTs. This duality also differs from the Pouliot-Strassler type duals~\cite{Pouliot:1995zc,Pouliot:1996zh}, which involve a symmetric tensor. It would be interesting to apply some of the techniques developed here to these chiral/nonchiral dualities with symmetric representations.

\section{Conclusions and future directions}\label{sec:concl}

In this work, we have studied the low energy dynamics and phase structure of chiral supersymmetric Yang-Mills theory with gauge group $SU(N)$, an antisymmetric tensor, and $F \le N+3$ flavors, in the presence of a cubic superpotential. We presented a rich set of dualities and phase transitions, together with novel nonperturbative and RG effects. Let us briefly summarize our results:
\begin{itemize}
\item[1)] For $F=N+3$ the theory flows to a superconformal fixed point. When $N$ is odd the theory admits a self-dual magnetic description. For arbitrary $N$, we found an infinite set of magnetic theories with gauge group $SU(N+K-1)$ (where $K$ is an arbitrary integer of the same parity as $N$), and a UV global symmetry group containing $SU(K)$. Fields charged under the additional magnetic global symmetry are eliminated from the chiral ring due to nonperturbative effects, and the $K$ dependence from the $a$-function and superconformal dimensions disappears at the IR fixed point.
\item[2)] For $F=N+2$ there are two different theories distinguished by the number of flavors involved in the cubic superpotential. One s-confines and has a weakly coupled description in terms of gauge singlets. The other flows to a superconformal fixed point, which admits a self-dual description (for even $N$), or another set of infinite magnetic duals similar to that in 1).
\item[3)] When $F=N+1$ the theory confines and breaks chiral symmetry. Thus the theory transitions from conformal to confining without an intermediate free magnetic phase.
\item[4)] If $F \le N$, a runaway instability develops, caused by nonperturbative effects. This is the analog of the ADS regime in SQCD, now in a chiral theory.
\end{itemize}

The behavior of the theory for $F > N+3$ seems to be more involved, containing more than one gauge group and leading to mixed phases of the type studied in~\cite{Terning:1997jj,Csaki:2004uj}. We leave the analysis of this very intriguing regime to future work \cite{appear}.

Our results also explain a puzzle found in works on deconfinement, first noticed in~\cite{Luty:1996cg}. Namely, the deconfinement method often leads to additional perturbative global symmetries, which are absent from the original theory. Consistency of the duality demands that for gauge-invariants in the chiral ring the global symmetries must agree, and in the past it was not always clear how fields charged under additional global symmetries would be removed. In this work, we found that new nonperturbative effects present in chiral theories are responsible for truncating matter charged under the additional global symmetries. We determined such effects using holomorphy and symmetries, as well as known instanton calculations for Sp gauge groups. It would be interesting to have a direct understanding of instanton effects in these chiral theories.

Decreasing the number of flavors by mass deformations, we found flows from superconformal to s-confining (between 1) and 2) above), and from s-confining or superconformal to confining with chiral symmetry breaking (between 2) and 3)).  Some of these transitions may prove fruitful in further understanding the phase structure of nonsupersymmetric QCD, where a transition from a conformal phase to a confining phase with chiral symmetry breaking is expected. Generalizing these models, it would be useful to systematically study the set of theories that lack a free magnetic phase.

\acknowledgments
We thank Daniel Green, Shamit Kachru and John Terning for useful discussions. NC is supported in part by the NSF under grant PHY-0907744 and gratefully acknowledges support from the Institute for Advanced Study.  
AH, RE and GT are supported by the US DOE under contract number DE-AC02-76SF00515. RE also acknowledges support by the National Science Foundation under Grant No.~NSF PHY05-51164.

\appendix

\section{Product gauge group flows for $F<N+3$}\label{sec:prodflow}

In this appendix, we construct a product gauge group that interpolates between the electric and magnetic theories with $F=N+2$, as a function of a holomorphic parameter.

Let us begin by integrating out $Q_{N+3}$ and $\tilde P$ for simplicity; this was the theory analyzed in \S \ref{subsec:tP}. Below this scale we therefore start from a product gauge group theory with matter content
\begin{center}
\vskip -6mm
\be\label{table:deconf-K2}
\begin{tabular}{c|cc|ccc}
&$SU(N)$&$Sp(N+K-4)$&$SU(K)$&$Sp(2N-2)$&$SU(N+2)$  \\
\hline
&&&&\\[-12pt]
$Q$&$\fund$&$1$&$1$&$1$&$\antifund$  \\
$\t Q$&$\antifund$&$1$&$1$&$ \fund$ &$1$ \\
$X$&$ \fund$&$\fund$&$ 1$&$1$ &$1$ \\
$U$&$\antifund$&$1$&$\antifund$&$ 1$ &$1$ \\
$V$&$1$&$\fund$&$\fund$&$ 1$ &$1$ \\
$T$&$1$&$1$&$\overline\antisym$&$1$ &$1$
\end{tabular}
\ee
\end{center}
The superpotential is, as before,
\be
W=  \t Q XX \t Q+XU V + VVT \,.
\ee
This theory can be studied in two different limits, depending on which gauge group factor becomes strong first.

If $\Lambda_{Sp(N+K-4)}\gg \Lambda_{SU(N)}$, the strong dynamics of the $Sp$ group dominates first, producing s-confinement. The theory in the infrared reduces precisely to (\ref{table:onelessflavor}), the electric theory with one fewer flavor. 

On the other hand, when $ \Lambda_{SU(N)}\gg \Lambda_{Sp(N+K-4)}$ the $SU(N)$ factor should be dualized first. We obtain
\begin{center}
\begin{tabular}{c|cc|ccc}
&$SU(N+K-2)$&$Sp(N+K-4)$&$SU(K)$&$Sp(2N-2)$&$SU(N+2)$  \\
\hline
&&&&\\[-12pt]
$q$&$\fund$&$1$&$1$&$1$&$\fund$  \\
$\t q$&$\antifund$&$1$&$1$&$ \fund$ &$1$ \\
$x$&$ \fund$&$\fund$&$ 1$&$1$ &$1$ \\
$u$&$\antifund$&$1$&$\fund$&$ 1$ &$1$ \\
$V$&$1$&$\fund$&$\fund$&$ 1$ &$1$ \\
$T$&$1$&$1$&$\overline\antisym$&$1$ &$1$ \\
$(Q \t Q)$&$1$&$1$&$1$&$\fund$ &$\antifund$ \\
$(Q  U)$&$1$&$1$&$\antifund$&$1$ &$\fund$ \\
$(X \t Q)$&$1$&$\fund$&$1$&$\fund$ &$1$ \\
$(X U)$&$1$&$\fund$&$\antifund$&$1$ &$1$
\end{tabular}
\end{center}
The superpotential now reads
\bea
W&=&  (\t Q X) (\t Q X)+(XU)V+VVT+\nonumber\\
&+&q (Q \t Q) \t q+q (Q U) u+x (X \t Q) \t q+x (X U) u\,.
\eea
The terms in the second line arise from Seiberg duality.

Integrating out the heavy fields leaves a {\it confining} $Sp(N+K-4)$ gauge group, which gives mesons $(xx)$ and breaks chiral symmetry. The confined theory superpotential including nonperturbative effects is
\bea
W&=& \t q (xx)\t q + (xx)uuT+ q (Q \t Q)\t q+ q (QU)u +\nonumber\\
&& + L\left(\text{Pf}(xx) - \Lambda_{Sp}^{N+K-2}\right)\,.
\eea
Here $L$ is a Lagrange multiplier field used to enforce the quantum modification of moduli space. The equations of motion for $L$ lead to a vacuum expectation value for $(xx)$, which higgses $SU(N+K-2) \to Sp(N+K-2)$ and gives mass to $(xx), \tilde q$. Integrating these fields out, the low energy theory then becomes
\begin{center}
\be\label{table:dualK-temp2}
\begin{tabular}{c|c|ccc}
&$Sp(N+K-2)$&$SU(K)$&$Sp(2N-2)$&$SU(N+2)$  \\
\hline
&&&\\[-12pt]
$q$&$\fund$&$1$&$1$&$ \fund$  \\
$u$&$\fund$&$\fund$&$ 1$ &$1$ \\
$(Q \t Q)$&$1$&$ 1$&$\fund$&$ \antifund$  \\
$(Q U)$&$ 1$&$\antifund$&$1$&$ \antifund$ \\
$T$&$1$&$\overline\antisym$&$1$ &$1$
\end{tabular}
\ee
\end{center}
with
\be\label{eq:WK2}
W=q (Q \t Q) (Q \t Q) q + uuT + q (Q U) u\,.
\ee

This theory is itself s-confining.  Upon confinement, integrating out massive fields $(qu), (uu), (QU), T$ yields 
\begin{center}
\be\label{table:dualK-temp3}
\begin{tabular}{c|cc}
&$Sp(2N-2)$&$SU(N+2)$  \\
\hline
&&\\[-12pt]
$(qq)$&$1$&$ \antisym$  \\
$(Q \t Q)$&$\fund$&$ \antifund$  \\
\end{tabular}
\ee
\end{center}
and superpotential $W = (Q \t Q)^2 (qq)$. After a renaming of fields, this again coincides precisely with the magnetic description of the dual in (\ref{table:mag-oneless}). This procedure also provides further evidence for the proposed duality of \cite{Spiridonov:2009za} by explicitly connecting the electric and magnetic theories in different limits of a holomorphic coupling.

The product gauge group theory for the case where one $\t Q$ is integrated out (discussed in \S \ref{subsec:tQ}) may be similarly constructed. The starting gauge group is again $SU(N) \times Sp(N+K-4)$. Dualizing $SU(N)$ first gives gauge groups $SU(N+K-2) \times Sp(N+K-4)$ and, upon integrating out the massive fields, we find that the $Sp$ factor s-confines. This generates a nonperturbative superpotential that has two types of terms; these reproduce the last two terms of the magnetic superpotential (\ref{eq:WmagN2tQ}).


\begin{thebibliography}{100}

\bibitem{Seiberg:1994bz}
  N.~Seiberg,
  Phys.\ Rev.\  {\bf D49}, 6857-6863 (1994).
  [hep-th/9402044].
N.~Seiberg,
  Nucl.\ Phys.\  {\bf B435}, 129-146 (1995).
  [hep-th/9411149].

\bibitem{Intriligator:1995au}
  K.~A.~Intriligator, N.~Seiberg,
  Nucl.\ Phys.\ Proc.\ Suppl.\  {\bf 45BC}, 1-28 (1996).
  [hep-th/9509066].

\bibitem{Affleck:1984uz}
  I.~Affleck, M.~Dine and N.~Seiberg,
  Phys.\ Rev.\ Lett.\  {\bf 52}, 1677 (1984).

\bibitem{Affleck:1984xz}
  I.~Affleck, M.~Dine and N.~Seiberg,
  Nucl.\ Phys.\  B {\bf 256}, 557 (1985).

\bibitem{Giudice:1998bp}
  G.~F.~Giudice, R.~Rattazzi,
  Phys.\ Rept.\  {\bf 322}, 419-499 (1999).
  [hep-ph/9801271].

\bibitem{Poppitz:1998vd}
  E.~Poppitz, S.~P.~Trivedi,
  Ann.\ Rev.\ Nucl.\ Part.\ Sci.\  {\bf 48}, 307-350 (1998).
  [hep-th/9803107].

\bibitem{Shadmi:1999jy}
  Y.~Shadmi, Y.~Shirman,
  Rev.\ Mod.\ Phys.\  {\bf 72}, 25-64 (2000).
  [hep-th/9907225].

\bibitem{Pouliot:1995zc}
  P.~Pouliot,
  Phys.\ Lett.\  {\bf B359}, 108-113 (1995).
  [hep-th/9507018].

\bibitem{Pouliot:1996zh}
  P.~Pouliot, M.~J.~Strassler,
  Phys.\ Lett.\  {\bf B375}, 175-180 (1996).
  [hep-th/9602031].
 P.~Pouliot, M.~J.~Strassler,
  Phys.\ Lett.\  {\bf B370}, 76-82 (1996).
  [hep-th/9510228].
T.~Kawano,
  Prog.\ Theor.\ Phys.\  {\bf 95}, 963-968 (1996).
  [hep-th/9602035].
M.~Berkooz, P.~L.~Cho, P.~Kraus, M.~J.~Strassler,
  Phys.\ Rev.\  {\bf D56}, 7166-7182 (1997).
  [hep-th/9705003].

\bibitem{Terning:1997jj}
  J.~Terning,
  Phys.\ Lett.\  B {\bf 422}, 149 (1998)
  [arXiv:hep-th/9712167].

\bibitem{Intriligator:1995ax}
  K.~A.~Intriligator, R.~G.~Leigh, M.~J.~Strassler,
  Nucl.\ Phys.\  {\bf B456}, 567-621 (1995).
  [hep-th/9506148].

\bibitem{Brodie:1996xm}
  J.~H.~Brodie, M.~J.~Strassler,
  Nucl.\ Phys.\  {\bf B524}, 224-250 (1998).
  [hep-th/9611197].

\bibitem{Berkooz:1995km}
  M.~Berkooz,
  Nucl.\ Phys.\  B {\bf 452}, 513 (1995)
  [arXiv:hep-th/9505067].

 \bibitem{Pouliot:1995me}
  P.~Pouliot,
  Phys.\ Lett.\  B {\bf 367}, 151 (1996)
  [arXiv:hep-th/9510148].

\bibitem{Csaki:2004uj}
  C.~Csaki, P.~Meade and J.~Terning,
  JHEP {\bf 0404}, 040 (2004)
  [arXiv:hep-th/0403062].

\bibitem{appear} N.~Craig, R.~Essig, A.~Hook, G.~Torroba, to appear.

\bibitem{Csaki:1997cu}
  C.~Csaki, M.~Schmaltz, W.~Skiba and J.~Terning,
  Phys.\ Rev.\  D {\bf 56}, 1228 (1997)
  [arXiv:hep-th/9701191].

\bibitem{Karch:1997jp}
  A.~Karch,
  Phys.\ Lett.\  {\bf B405}, 280-286 (1997).
  [hep-th/9702179].

\bibitem{Intriligator:2003jj}
  K.~A.~Intriligator and B.~Wecht,
  Nucl.\ Phys.\  B {\bf 667}, 183 (2003)
  [arXiv:hep-th/0304128].

\bibitem{Luty:1996cg}
  M.~A.~Luty, M.~Schmaltz and J.~Terning,
  Phys.\ Rev.\  D {\bf 54}, 7815 (1996)
  [arXiv:hep-th/9603034].


\bibitem{Leigh:1996ds}
  R.~G.~Leigh, M.~J.~Strassler,
  Nucl.\ Phys.\  {\bf B496}, 132-148 (1997).
  [hep-th/9611020].

\bibitem{Distler:1996ub}
  J.~Distler, A.~Karch,
  Fortsch.\ Phys.\  {\bf 45}, 517-533 (1997).
  [hep-th/9611088].



\bibitem{Poppitz:1995fh}
  E.~Poppitz and S.~P.~Trivedi,
  Phys.\ Lett.\  B {\bf 365}, 125 (1996)
  [arXiv:hep-th/9507169].

\bibitem{Intriligator:1995ne}
  K.~A.~Intriligator and P.~Pouliot,
  Phys.\ Lett.\  B {\bf 353}, 471 (1995)
  [arXiv:hep-th/9505006].



\bibitem{Kutasov:2003iy}
  D.~Kutasov, A.~Parnachev and D.~A.~Sahakyan,
  JHEP {\bf 0311}, 013 (2003)
  [arXiv:hep-th/0308071].

\bibitem{Barnes:2004jj}
  E.~Barnes, K.~A.~Intriligator, B.~Wecht and J.~Wright,
  Nucl.\ Phys.\  B {\bf 702}, 131 (2004)
  [arXiv:hep-th/0408156].

\bibitem{Spiridonov:2009za}
  V.~P.~Spiridonov, G.~S.~Vartanov,
  Commun.\ Math.\ Phys.\  {\bf 304}, 797-874 (2011).
  [arXiv:0910.5944 [hep-th]].
  
\bibitem{Romelsberger:2007ec}
  C.~Romelsberger,
  [arXiv:0707.3702 [hep-th]].

\bibitem{Romelsberger:2005eg}
  C.~Romelsberger,
  Nucl.\ Phys.\  {\bf B747}, 329-353 (2006).
  [hep-th/0510060].


  
  \bibitem{Bailey}
  V.~P.~Spiridonov,
  Teor.\ Mat.\ Fiz.\ {\bf 139} (2004).
  [math.CA/0312502].  V.~P.~Spiridonov, and S.~O.~Warnaar,
  Adv.\ Math.\  {\bf 207} (2006).


\end{thebibliography}
\end{document}